\documentclass[showpacs,aps,prb,reprint,superscriptaddress,preprintnumbers]{revtex4-2}

\usepackage{amsmath}
\usepackage{amssymb}
\usepackage{graphicx}
\usepackage{dcolumn}
\usepackage[dvipsnames]{xcolor}
\usepackage{amsthm,amssymb}
\usepackage{mathrsfs}

\usepackage{color}
\usepackage{endnotes}
\usepackage{bm}
\usepackage{epsfig}
\usepackage{array}
\usepackage{ulem}
\usepackage{soul}

\newcommand{\CsVSb}{{CsV$_{3}$Sb$_{5}$}}
 
\newcommand{\parallelsum}{\mathbin{\!/\mkern-5mu/\!}}

\begin{document}

\title{Large Fermi surface in pristine kagome metal CsV$_3$Sb$_5$ and \\ enhanced quasiparticle effective masses}

\author{Wei~Zhang}
\author{Tsz~Fung~Poon}
\author{Chun Wai Tsang}
\author{Wenyan~Wang}
\author{X.~Liu}
\author{J.~Xie}
\author{S. T. Lam}
\affiliation{Department of Physics, The Chinese University of Hong Kong, Shatin, Hong Kong, China}
\author{Shanmin~Wang}
\affiliation{Department of Physics, Southern University of Science and Technology, Shenzhen, Guangdong, China}
\author{Kwing~To~Lai}
\affiliation{Department of Physics, The Chinese University of Hong Kong, Shatin, Hong Kong, China} 
\affiliation{Shenzhen Research Institute, The Chinese University of Hong Kong, Shatin, Hong Kong, China}
\author{A. Pourret}
\affiliation{Univ. Grenoble Alpes, CEA, Grenoble-INP, IRIG, Pheliqs, 38000 Grenoble, France}
\author{G. Seyfarth}
\affiliation{Univ. Grenoble Alpes, INSA Toulouse, Univ. Toulouse Paul Sabatier,
EMFL, CNRS, LNCMI, Grenoble 38042, France}
\author{G. Knebel}
\email[]{georg.knebel@cea.fr}
\affiliation{Univ. Grenoble Alpes, CEA, Grenoble-INP, IRIG, Pheliqs, 38000 Grenoble, France}
\author{Wing~Chi~Yu}
\email[]{wingcyu@cityu.edu.hk}
\affiliation{Department of Physics, City University of Hong Kong, Kowloon, Hong Kong, China}
\author{Swee~K.~Goh}
\email[]{skgoh@cuhk.edu.hk}
\affiliation{Department of Physics, The Chinese University of Hong Kong, Shatin, Hong Kong, China}

\date{\today}

\begin{abstract}

{The kagome metal CsV$_3$Sb$_5$ is an ideal platform to study the interplay between topology and electron correlation. To understand the fermiology of CsV$_3$Sb$_5$, intensive quantum oscillation (QO) studies at ambient pressure have been conducted. However, due to the Fermi surface reconstruction by the complicated charge density wave (CDW) order, the QO spectrum is exceedingly complex, hindering a complete understanding of the fermiology. Here, we directly map the Fermi surface of the pristine CsV$_3$Sb$_5$ by measuring Shubnikov-de Haas QOs up to 29~T under pressure, where the CDW order is completely suppressed. The QO spectrum of the pristine \CsVSb\ is significantly simpler than the one in the CDW phase, and the detected oscillation frequencies agree well with our density functional theory calculations. In particular, a frequency as large as 8,200~T is detected. Pressure-dependent QO studies further reveal a weak but noticeable enhancement of the quasiparticle effective masses on approaching the critical pressure where the CDW order disappears, hinting at the presence of quantum fluctuations. Our high-pressure QO results reveal the large, unreconstructed Fermi surface of CsV$_3$Sb$_5$, paving the way to understanding the parent state of this intriguing metal in which the electrons can be organized into different ordered states.}

\end{abstract}

\maketitle

Kagome metals AV$_3$Sb$_5$ (A=K, Rb, Cs), whose electronic band structure naturally hosts van Hove singularities, Dirac points and flat bands, serve as an ideal platform to investigate the interplay between topologically nontrivial states and the electronic correlation. Due to the correlation effects, an abundance of instabilities, including charge density wave (CDW), nematicity and superconductivity (SC), have been realized in AV$_3$Sb$_5$~\cite{Ortiz2019,Ortiz2020,Ortiz2021,Kiesel2012,Kiesel2013,Wang2013,Nie2022,Li2022,Xu2022,Yin2021,Tan2021,Wang2023,Wang2023b,Yang2020,Du2021,Wang2021a,Zheng2022,Kang2022,Feng2023,Asaba2023,Tan2023}. Hence, AV$_3$Sb$_5$ metals exhibit various quantum states that are actively explored in modern condensed matter research.  

Among the three members of AV$_3$Sb$_5$, \CsVSb\ possesses the highest superconducting transition temperature ($T_{\rm c} \sim$ 2.7~K) with a CDW transition at $T_{\rm CDW} \sim$ 90~K~\cite{Ortiz2019,Ortiz2020,Du2021,Wang2021a}. The precise nature of the CDW order in \CsVSb\ is currently under intense scrutiny, and contradictory results have been reported. For instance, whether time-reversal symmetry is preserved in the CDW phase is an important question. While muon spin relaxation, chiral transport measurements, an earlier magneto-optical polar Kerr effect measurement, and the detection of the anomalous Hall effect ~\cite{Khasanov2022,Guo2022,Hu2022,Yu2021c,Xu2022,Yu2021b} support the absence of the time-reversal symmetry, the most recent magneto-optical Kerr result indicates the preservation of the time-reversal symmetry~\cite{Saykin2023,Farhang2023}.
Next, the nature of the superlattice modulation is also under debate~\cite{Ortiz2020,Ortiz2021,Li2021,Liang2021,Kang2023a}. It has been revealed that the in-plane distortion can adopt a $2\times 2$ superlattice with Star of David (SoD) or trihexagonal (TrH) patterns. However, the out-of-plane distortion mode which involves various possible stacking of SoD and TrH patterns remains elusive, and both $2\times 2\times 2$ and $2\times 2\times 4$ modulation modes have been reported~\cite{Ortiz2020,Ortiz2021,Li2021,Liang2021,Hu2022a,Kang2023a}. 

The emergence of the new periodicity associated with the CDW order introduces a smaller Brillouin zone (BZ). Thus, the pristine Fermi surface is reconstructed because of the band folding, and an additional energy gap develops which can affect the band structure. Due to the BZ folding and Fermi surface reconstruction, many small Fermi pockets can be expected. The CDW-induced energy gap has been detected by scanning tunneling microscopy and angle-resolved photoemission spectroscopy~\cite{Liang2021,Hu2022a,Kang2023a}.
Hence, knowing the superlattice distortion pattern in the CDW phase is essential for calculating the electronic band structure. Intensive quantum oscillation (QO) studies have revealed an abundance of frequencies in \CsVSb\ at ambient pressure~\cite{Ortiz2021,Yu2021,Fu2021,Yu2021b,Gan2021,Zhang2022,Chen2022,Huang2022,Shrestha2022,Broyles2022,Chapai2023}. However, due to the limited knowledge of the CDW superlattice distortion pattern, a good consistency of the fermiology between experiments and density functional theory (DFT) calculations has not been achieved.

To make progress towards understanding the fermiology of \CsVSb, a natural strategy is to first remove the complicated CDW order, and then probe the pristine metallic phase by QOs.  This can be accomplished by turning to the well-known temperature-pressure ($T$-$p$) phase diagram, which shows that $T_{\rm CDW}$ decreases monotonically with increasing pressure and extrapolates to 0~K at $p_c$ $\sim$ 20 kbar~\cite{Yu2021,Chen2021a}. Furthermore, the $T$-$p$ phase diagram indicates an interesting interplay between the CDW order and SC -- the SC displays a double-dome pressure-dependence with a maximum $T_{\rm c}$ at $p_c$. While the origin of the first dome near 7~kbar is not yet settled, the second dome around $p_c$ resembles the phenomenon of superconductivity mediated by quantum criticality~\cite{Feng2023, Tazai2022}, where quantum fluctuations associated with the CDW instability can enhance superconductivity analogous to various correlated electron systems, including heavy fermion systems, cuprates and iron-based superconductors~\cite{Shishido2005,Shishido2010,Ramshaw2015}. Therefore, the pressure evolution of fermiology around $p_c$ can shed light on the presence of quantum fluctuations.  

In this study, we measure Shubnikov-de Haas QOs of \CsVSb\ under pressure up to 29~T. We investigate the Fermi surface of the pristine phase, and explore the pressure dependence of the fermiology. In the pristine \CsVSb\ with the magnetic field ($B$) along the $c$-axis, a QO frequency as high as $\sim$8,250~T is resolved in our study. Indeed, the entire fast Fourier transform (FFT) spectrum of the pristine phase is significantly simpler, and it shows a much-improved agreement with DFT calculations. Finally, we uncover a weak but discernible enhancement of the quasiparticle effective masses upon approaching $p_c$ on the pristine side. Thus, a near-complete determination of the Fermi surface sets the scene for understanding the parent metallic state which is susceptible to various instabilities, and the pressure-dependent fermiology parameters shed light on the effect of fluctuations associated with these instabilities on conduction electrons.
\\
 
\begin{figure}[!t]\centering
      \resizebox{8.8cm}{!}{
              \includegraphics{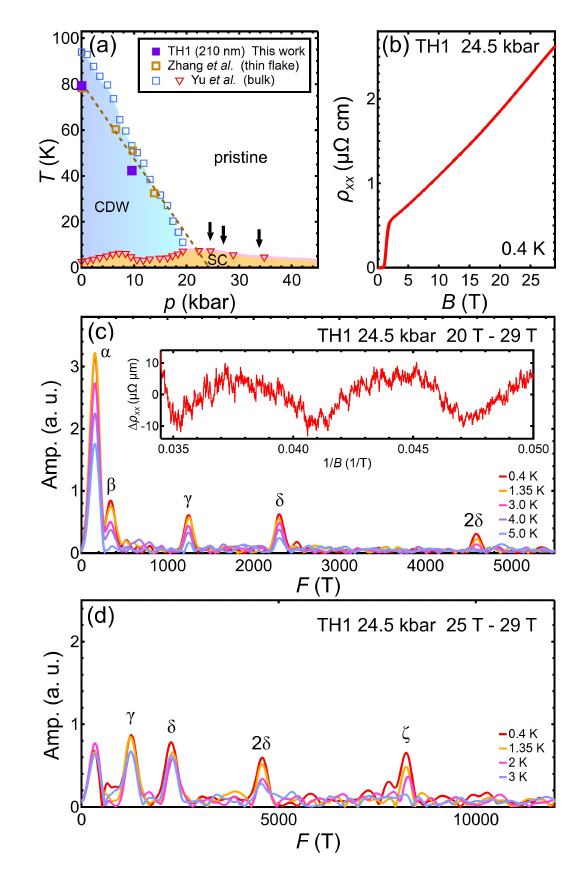}}                				
              \caption{\label{fig1}  
              (a) Temperature-pressure phase diagram of \CsVSb. The open symbols for bulk and thin flake are adapted from Ref.~\cite{Yu2021} and Ref.~\cite{Zhang2023}, respectively. The brown dashed line is the linear fit of the $T_{\rm CDW}$ for thin flake of Ref.~\cite{Zhang2023}, which gives 24 kbar for the CDW suppression point. The black arrows point to 24.5, 27.0 and 33.8 kbar, indicating the pressure values chosen for the QO studies in the pristine phase of this work. (b) $\rho_{xx}(B)$ at 0.4 K. (c) FFT spectra of the thin flake TH1 for the oscillatory data between 20~T and 29~T at different temperatures. The inset shows the oscillatory signals after the removal of the background at 0.4 K. (d) FFT spectrum of TH1 for the data between 25~T and 29~T at different temperatures.}
\end{figure}

\begin{figure*}[!t]\centering
      \resizebox{18cm}{!}{
 \includegraphics{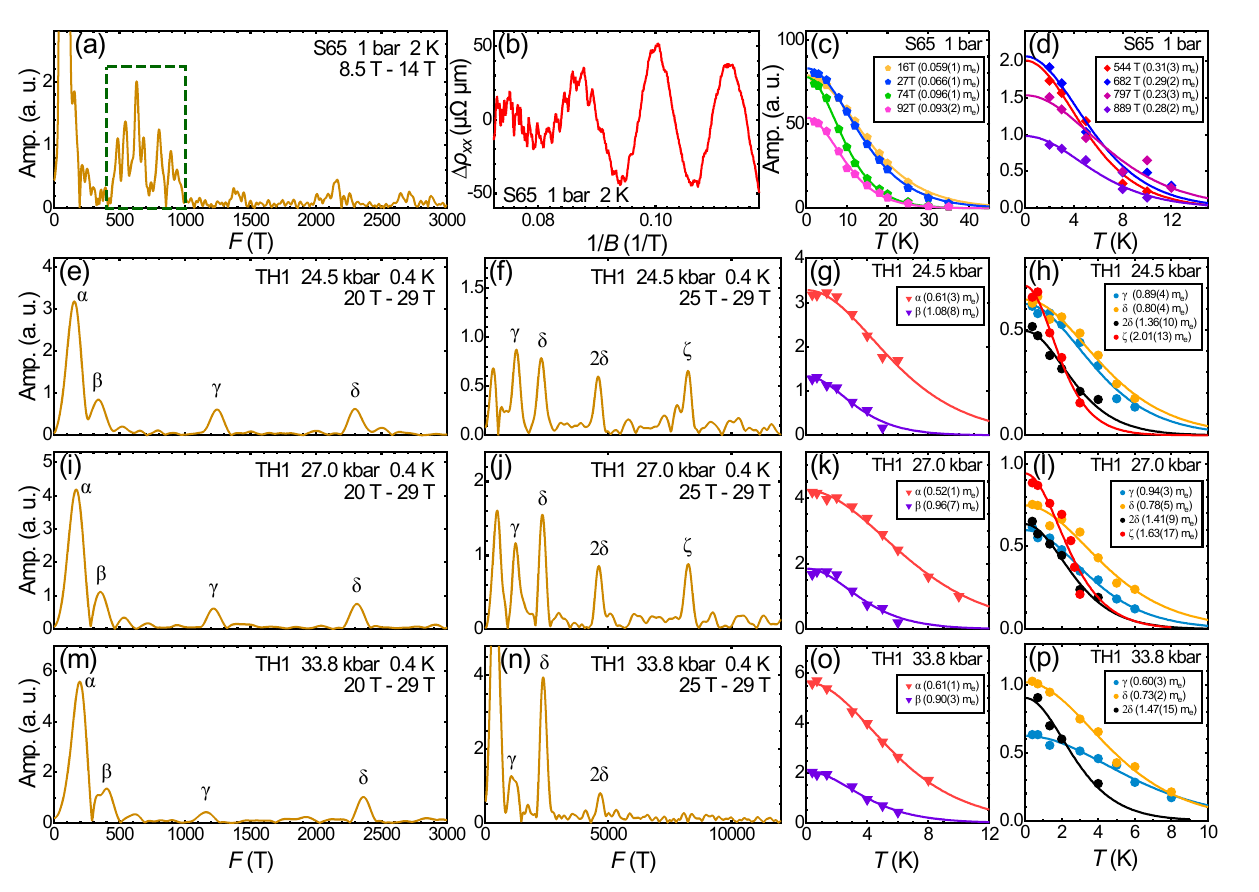}}       \caption{\label{fig2} 
(a) FFT spectrum of the thin flake S65 for the data from 8.5~T to 14~T at ambient pressure. (b) Oscillatory signals after the removal of the background for S65. (c), (d) Temperature dependence of QO amplitudes for S65 at ambient pressure. FFT spectra of TH1 at 0.4 K for (e), (f) 24.5 kbar, (i), (j) 27.0 kbar and (m), (n) 33.8 kbar. Temperature dependence of QO amplitudes of TH1 for (g), (h) 24.5 kbar, (k), (l) 27.0 kbar and (o), (p) 33.8 kbar. The solid
curves are fits using the thermal damping factor $R_T$ of Lifshitz-Kosevich theory.  }          
\end{figure*}

\noindent{\bf \large{Results}}

\noindent{\bf Large Fermi surface in the pristine phase}

\noindent 
Figure~\ref{fig1}(a) shows the $T$-$p$ phase diagram of \CsVSb. For the bulk sample, the CDW order can be totally suppressed at 20 kbar and for the thin flakes, the CDW order disappears at a slightly higher pressure of 24 kbar~\cite{Yu2021,Chen2021a,Zhang2023}. As reported by some of us in Ref.~\cite{Zhang2023}, above 24 kbar, both the anomaly feature in ${\rm d}\rho{\rm /}{\rm d}T$ and the characteristic ``S-shape" line in the low field region of $\rho_{yx}(B)$ disappear, indicating the total suppression of the CDW order. In this work, we measure the QO in a thin flake of \CsVSb\ with a thickness of 210~nm (TH1). As displayed in Fig.~\ref{fig1}(a), the $T_{\rm CDW}$ of TH1 at ambient pressure and 9.6~kbar agrees well with the trend of $T_{\rm CDW}(p)$ for the thin flakes (brown dashed line). Hence, we choose 24.5~kbar as the first pressure point to study the pristine phase. As shown in Fig.~S1, $\rho_{yx}(B)$ at 24.5~kbar changes slowly without the characteristic ``S-shape" line at the low-field region, again, indicating the removal of the CDW order. We then conduct the magnetotransport measurement and Fig.~\ref{fig1}(b) shows the longitudinal resistivity ($\rho_{xx}$) up to 29~T. The signal quality is superior, with the superconducting to normal-state transition clearly observed. A low extrapolated normal-state $\rho_{xx}({\rm 0~T, 0.4~K})$ of 0.43~$\mu\Omega{\rm cm}$, representative of the residual resistivity, indicates a high-quality sample. Using the extrapolated $\rho_{xx}({\rm 0~T, 0.4~K})$, the magnetoresistance is calculated to reach 510\% and it is non-saturating at 29~T. It is interesting to note that the normal-state $\rho_{xx}$ is quasilinear in $B$, consistent with a previous report~\cite{Yu2021b}. To search for QOs in $\rho_{xx}$, we perform a slow field sweep between 20~T and 29~T. After the removal of the background, excellent QO signals show up as displayed in the inset of Fig.~\ref{fig1}(c). The FFT spectrum for the data between 20~T and 29~T gives five peaks, namely $\alpha$ (154(62)~T), $\beta$ (343(63)~T), $\gamma$ (1,244(56)~T), $\delta$ (2,298(57)~T) and $2\delta$ (4,589(57)~T). The 4,589~T peak is assigned as 2$\delta$ not only because the frequency is double that of $\delta$ at all pressures, but its quasiparticle effective mass ($m^*$) is also nearly twice that of $\delta$ (see the effective mass analysis below and Table S1). If the error bars are considered, it is also possible that $\beta$ is a harmonic of $\alpha$ (see Table S2 for more details). Besides, a careful analysis of the high-field data above 25~T reveals the existence of another peak with a frequency of 8,235(165)~T ($\zeta$), as shown in Fig.~\ref{fig1}(d). The detection of such a large frequency is significant, as this frequency translates to an orbit larger than the area of the first BZ in the $k_x$-$k_y$ plane when a 2$\times$2 in-plane superlattice modulation is considered ($\sim$4,000~T, using $a$=5.4651~\AA\ at 30~kbar~\cite{Tsirlin2022}). Thus, the detection of $\zeta$ is consistent with the absence of a CDW order. For the pristine \CsVSb, the area of the first BZ corresponds to a frequency of $\sim$16,000~T (calculated by taking $a$=5.4651~\AA\ at 30~kbar~\cite{Tsirlin2022}). Hence, the resolved $\zeta$ peak occupies $\sim$51\% of the first BZ area. Overall, we successfully resolved six QO frequencies in the pristine phase of \CsVSb.

The FFT spectrum at 24.5 kbar is much simpler compared with the spectrum at ambient pressure~\cite{Zhang2022}.
Figure~\ref{fig2}(a) shows the FFT spectrum of \CsVSb\ (S65) at ambient pressure and the corresponding oscillatory signals after the removal of the background is shown in Fig.~\ref{fig2}(b). The most obvious difference between the ambient pressure and the pristine data is that the abundance of peaks with fine structure between 400~T and 1,000~T at ambient pressure (emphasized by the green rectangle) disappear in the pristine case. In fact, these frequencies at ambient pressure have been revealed by different techniques, such as tunnel diode oscillator frequency measurements up to 41.5~T~\cite{Broyles2022}. The overall FFT spectrum for S65 is also consistent with our ambient pressure study before~\cite{Zhang2022}. It might be tempting to attribute the disappearance of the peaks between 400~T and 1,000~T in the pristine phase to a possible enhancement of $m^*$ for these frequencies. However, we are able to detect much larger frequencies with heavier $m^*$ ({\it e.g.} $\zeta$ with an effective mass as high as $\sim$2~$m_e$, as shown in Figs.~\ref{fig2}(f) and ~\ref{fig2}(h).), indicating the superior signal-to-noise ratio of our experiment. Thus, the disappearance of these frequencies in the pristine phase is most likely attributable to the absence of Fermi surface reconstruction, introduced because of the additional periodicity associated with the onset of the CDW transition. The BZ folding and Fermi surface reconstruction could induce many small Fermi pockets (hence small QO frequencies). When the CDW order is suppressed, such reconstruction does not occur, resulting in a simpler QO spectrum.

Next, we move further away from the CDW phase by increasing the applied pressure. The FFT spectra at 27.0~kbar and 33.8~kbar are displayed in Figs.~\ref{fig2}~(i)-(j) and Figs.~\ref{fig2}~(m)-(n), respectively. These spectra closely resemble the FFT spectrum at 24.5~kbar. With the same experimental parameters, all peaks can be consistently followed from 24.5~kbar to 33.8~kbar, except for $\zeta$. However, when the field sweeping rate is slower and the excitation current is higher, $\zeta$ does show up at 33.8~kbar but with a significantly reduced amplitude (see Fig.~S2).  Finally, the group of peaks between 400~T and 1,000~T is absent, in agreement with the fact that all these pressure points are located beyond the CDW phase. Therefore, we conclude that the disappearance of the group of peaks and the emergence of $\zeta$ are fingerprints of the pristine metallic state of \CsVSb.

Oscillation peaks with frequencies larger than the in-plane area of the reconstructed BZ have been reported in Ref.~\cite{Chapai2023} and they are attributed to magnetic breakdown. The situation is dissimilar to our case because, in Ref.~\cite{Chapai2023}, the high-frequency peaks are observed in the CDW phase while $\zeta$ in our study is seen in the pristine phase. Furthermore, the high-frequency peaks in Ref.~\cite{Chapai2023} are equally spaced with an interval of $\sim$2000~T, clearly different from our FFT spectra with a single, well-defined $\zeta$.  
Next, the magnetic field used in Ref.~\cite{Chapai2023} is 86~T, while our maximum field is 29~T. Finally, as will be discussed in the later section, our data show a good agreement with the calculated Fermi surface in the pristine phase. Therefore, we adopt the most natural explanation instead of invoking the notion of a magnetic breakdown.
\\

\begin{figure}[!t]\centering
       \resizebox{8.5cm}{!}{
     \includegraphics{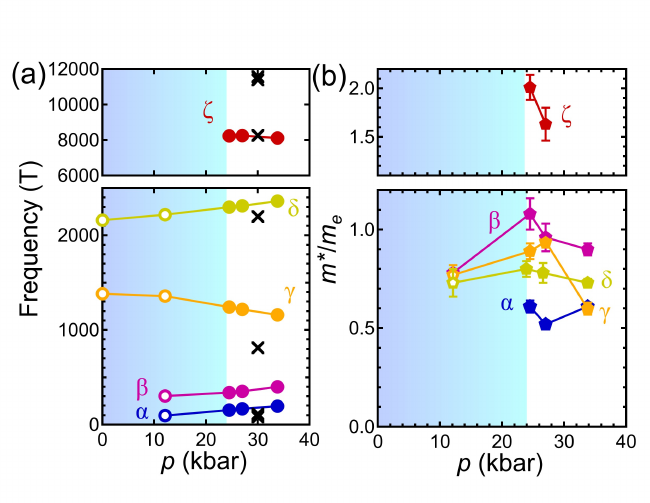}}         
              \caption{\label{fig3}Pressure-evolution of (a) quantum oscillation frequencies and (b) quasiparticle effective masses. The black crosses in (a) represent the calculated QO frequencies at 30 kbar. The coloured region denotes the CDW phase. The frequency of $\zeta$ at 33.8 kbar is extracted from Fig.~S2, in which a different field range is used for the FFT.}
\end{figure}

\noindent{\bf Pressure dependent quasiparticle effective masses}

\noindent The temperature dependencies of the QO amplitudes at 24.5 kbar are shown in Figs.~\ref{fig2}(g)-(h) and analyzed using the thermal damping factor ($R_T$) of Lifshitz-Kosevich theory. The analyses give $m^*/m_e$ of 0.61(3) ($\alpha$), 1.08(8) ($\beta$), 0.89(4) ($\gamma$), 0.80(4) ($\delta$), 1.36(10) (2$\delta$) and 2.01(13) ($\zeta$). First, within the uncertainties, we note that $m^*$ of $2\delta$ is nearly twice the value for $\delta$ (see the effective mass analysis below and Table S1). Second, the $m^*$ values at 24.5 kbar are noticeably larger than the values at ambient pressure, which are all less than $\sim$0.3~$m_e$ (see Figs.~\ref{fig2}(c)-(d)). In particular, $m^*$ of $\zeta$ is as high as 2.01(13)~$m_e$. Similarly, the QO amplitudes at 27.0~kbar and 33.8~kbar follow the Lifshitz-Kosevich theory well (Figs.~\ref{fig2}(k)-(l) and Figs.~\ref{fig2}(o)-(p)), enabling the calculation of $m^*$. The $m^*$ values at 27.0~kbar and 33.8~kbar are also markedly larger than the overall values at ambient pressure, and the ratio of $m^*(2\delta)$ to $m^*(\delta)$ is found to be 2 as well, as tabulated in Table S1.

The observation of QO frequencies over a wide pressure range offers an ideal backdrop to discuss the evolution of $m^*$. Figure~\ref{fig3}(a) plots the QO frequencies as a function of pressure, where the solid symbols are values for the pristine phase extracted from the spectra displayed in Fig.~S2 (for $\zeta$ at 33.8~kbar) and Fig.~\ref{fig2} (for all other datapoints). Selected QO frequencies from the CDW phase that show an obvious connection to the frequencies in the pristine phase are also included as open symbols in Fig.~\ref{fig3}(a). The FFT spectra for TH1 at 12.1~kbar are shown in Fig. S3.
Thus, $\gamma$ and $\delta$ can be seen from ambient pressure to 33.8~kbar, and they vary roughly linearly over the full pressure range. Figure~\ref{fig3}(b) shows the pressure dependence of $m^*$ from 12.1~kbar to 33.8~kbar. At 12.1~kbar, we can determine the $m^*$ of $\beta$, $\gamma$ and $\delta$ because of the less complicated FFT spectrum (Fig.~S3). However, it is challenging to do the same for the ambient pressure data because oscillation peaks have a fine structure and the weak signal-to-noise ratio for $\gamma$ and $\delta$ prohibit an accurate mass analysis. 

\begin{figure*}[!t]\centering
      \resizebox{12cm}{!}{
 \includegraphics{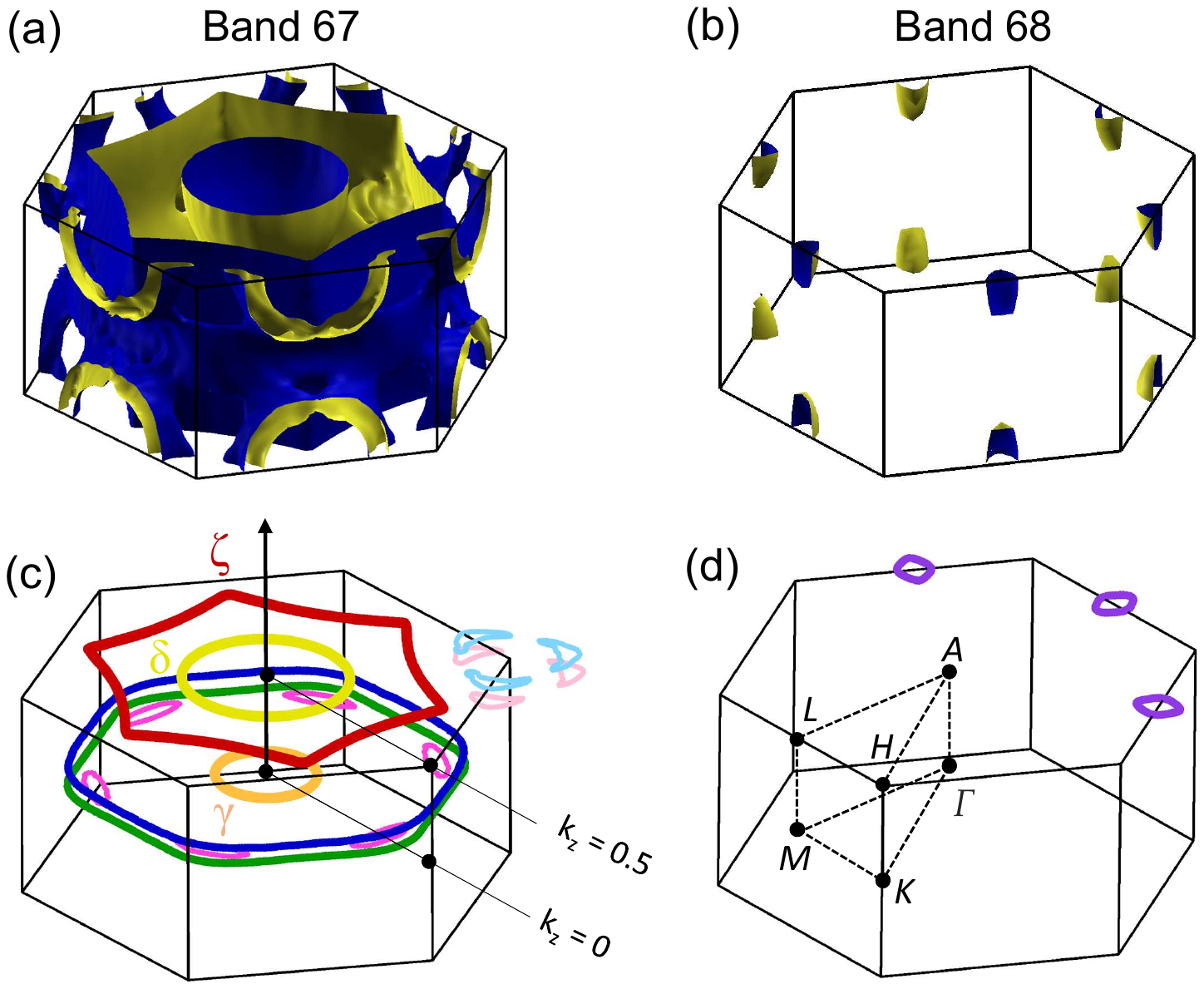}}       \caption{\label{fig4}  
Calculated Fermi surface of the pristine \CsVSb\ at 30 kbar for (a) Band~67 and (b) Band~68. The extracted extremal orbits for Band~67 and Band~68 are shown in (c) and (d), respectively. For clarity, some duplicated copies of the extremal orbits are not shown. See text for details.}           
\end{figure*}

The effective masses presented in Fig.~\ref{fig3}(b) all come from the same sample (TH1) measured at the same high-field facility. Therefore, their pressure evolution is particularly meaningful. Near the pressure where $T_{\rm CDW}$ vanishes, $m^*$ of $\zeta$, $\beta$ and $\gamma$ show a noticeable maximum, while $m^*$ of $\delta$ is roughly constant when the error bars are considered. Calculations performed at 30~kbar and 49~kbar, namely on the pristine side, show that the calculated band structures near $E_F$ are nearly identical, except for the band near $\Gamma$ (see Fig.~S5), which becomes flatter at 49~kbar. This means the band masses ($m^*_{cal}$) are almost constant in this pressure range, except for the band mass associated with the band near $\Gamma$, which would be larger at 49~kbar than at 30~kbar. Thus, the ratio of the experimental effective mass to the calculated band mass ($m^*/m^*_{cal}$) of $\beta$, $\gamma$, and $\delta$ increase when approaching the CDW phase boundary on the pristine side. A similar analysis cannot be extended into the CDW phase because a full knowledge of the superlattice distortion pattern is needed to calculate the band masses.

In heavy fermion CeRhIn$_5$ and iron-based superconductor BaFe$_2$(As,P)$_2$, quantum oscillation data showed the enhancement of $m^*$ when they are tuned towards the magnetic quantum critical point~\cite{Shishido2005,Shishido2010}. In addition, $m^*$ is also observed to increase significantly in YBa$_2$Cu$_3$O$_{6+\delta}$ tuned towards the critical hole concentration of $p=0.18$, where the role of a possible CDW fluctuations have been put forward~\cite{Ramshaw2015}. The possibility of a quantum critical point in CsV$_3$Sb$_5$ when $T_{\rm CDW}\rightarrow0$ has been pointed out by theoretical studies and high-pressure nuclear quadrupole resonance data~\cite{Tazai2022,Wang2022,Feng2023}. Furthermore, the initial slope of the upper critical field, $(-dH_{c2}/dT)_{T_c}$, exhibits a maximum at the pressure where $T_{\rm CDW}\rightarrow0$~\cite{Chen2021a}. Our QO studies do not give strongly diverging effective masses in CsV$_3$Sb$_5$. However, fluctuations related to the suppression of the CDW may still play a role in the pressure phase diagram, which can partially explain the enhanced, but non-diverging, quasiparticle effective masses near 24~kbar.

\noindent{\bf\\ Near-complete determination of the Fermi surface}

\noindent To understand the fermiology of \CsVSb\ in the pristine phase, we perform DFT calculations. Both lattice parameters and atomic positions are fixed to experimental values at 30~kbar provided by Ref.~\cite{Tsirlin2022}. Our DFT calculations show that only two bands cross the Fermi level in the pristine phase at 30 kbar, and the calculated Fermi surfaces are displayed in Figs.~\ref{fig4}(a) and (b). Meanwhile, as shown in Table~\ref{table1}, the two bands give 9 QO frequencies when the magnetic field is along the $c$-axis. We find that the calculated Fermi surface does not host a series of QO frequencies with a fine structure between 400 and 1,000~T. Instead, the calculation only predicts a single, well-defined frequency at 812~T, which is consistent with our experimental results in the pristine phase (Fig.~\ref{fig2}).

\begin{table*}[t!]
\centering
\caption{\label{table1}Comparison of the calculated quantum oscillation results (30.0 kbar) and the detected SdH quantum oscillation data (27.0 kbar and 33.8 kbar) with $B\,\parallelsum \,c$. The frequency of $\zeta$ at 33.8 kbar is extracted from Fig.~S2, in which a different field range is used for the FFT.}
\begin{tabular}{p{1.5cm}  p{1.5cm} | p{1.5cm}  p{1.5cm}  p{1.5cm}  p{1.5cm} | p{1.5cm}  p{1.5cm}}
\hline
\hline
 \multicolumn{2}{c|}{27.0 kbar (Experimental data)}  & \multicolumn{4}{c|}{30.0 kbar (DFT)}  & \multicolumn{2}{c}{33.8 kbar (Experimental data)} \\  
 \hline
 \centering $F$ (T)  & \centering $m^*/m_e$ & &\centering $k_z$ ($2\pi/c$)  &\centering $F_{cal}$ (T)    &\centering $m^*_{cal}/m_e$ &\centering $F$ (T)  & $m^*/m_e$\\ 
\hline

\centering $-$ &\centering $-$ & \centering Band 67  &\centering $\pm$0.41 &\centering  82 &\centering  0.3  &\centering $-$ &\qquad $-$  \\     

\centering $-$ &\centering $-$ & &\centering 0.50 &\centering  96 &\centering  0.3  &\centering $-$ &\qquad $-$ \\     

\centering $-$ &\centering $-$ & &\centering 0 &\centering  103 &\centering  0.3 &\centering $-$ &\qquad $-$ \\     

\centering 1,217 ($\gamma$) &\centering 0.94(3) & &\centering 0 &\centering  812 &\centering  0.5  &\centering 1,161 ($\gamma$) &\ \quad 0.60(3)\\     

\centering 2,313 ($\delta$) &\centering 0.78(5) & &\centering 0.50 &\centering  2,199 &\centering  0.4 &\centering 2,363 ($\delta$) &\ \quad 0.73(2)\\     

\centering 8,243 ($\zeta$) &\centering 1.63(17) & &\centering 0.50 &\centering  8,266 &\centering  1.1  &\centering 8,125 ($\zeta$) &\qquad $-$\\
  
 \centering $-$ &\centering $-$ & &\centering 0 &\centering  11,404 &\centering  3.5  &\centering $-$ &\qquad $-$  \\     

 \centering $-$ &\centering $-$ & &\centering $\pm$0.08 &\centering  11,577 &\centering  5.1  &\centering $-$ &\qquad $-$ \\     
\hline
 \centering 168 ($\alpha$) &\centering 0.52(1) & \centering $-$  &\centering $-$ &\centering  $-$ &\centering  $-$  &\centering 195 ($\alpha$) &\ \quad 0.61(1)\\     
\hline
 \centering 354 ($\beta$) &\centering 0.96(7)  &\centering $-$  &\centering  $-$&\centering $-$ &\centering  $-$  &\centering 401 ($\beta$) &\ \quad 0.90(3)\\     
\hline
 \centering $-$ &\centering $-$ & \centering Band 68  &\centering 0.50 &\centering  112 &\centering  0.1  &\centering $-$ &\ \quad $-$\\     
\hline
\end{tabular}
\end{table*}

To further compare the calculated Fermi surface with our QO data, we add the calculated frequencies on the frequency-pressure diagram (Fig.~\ref{fig3}(a)). The calculated frequencies agree with our experimental results well. The $\alpha$ and $\beta$ frequencies correspond to the calculated low frequencies below 400~T, and $\delta$, $\zeta$ can be associated to the two frequencies at 2,199~T and 8,266~T, respectively. We note that there are also calculated frequencies above 10,000~T, which are not captured in our QO study. Considering the relatively large calculated band masses (Table~\ref{table1}), further study with a stronger magnetic field or purer crystals is needed to detect these high-frequency oscillations.  There is one calculated frequency (812~T) left, which can be tied to $\gamma$ because $\gamma$ is the remaining unassigned frequency closest to 812~T, and the observed decreasing trend under pressure is consistent with calculations (see further discussion below)~\cite{Tsirlin2022,Wenzel2023}. Hence, our QO measurements in the pristine phase successfully capture nearly all frequencies from the DFT calculations at 30~kbar. This is a very different scenario from ambient-pressure studies, where achieving a satisfactory agreement between the experimental results and calculated frequencies has been challenging, likely due to the complex CDW order where the nature of the superlattice distortion is still under debate~\cite{Ortiz2021,Yu2021,Fu2021,Yu2021b,Gan2021,Zhang2022,Chen2022,Huang2022,Shrestha2022,Broyles2022,Chapai2023}. Therefore, the QO dataset in the pristine phase represents a key step towards a complete experimental determination of the fermiology of \CsVSb.

The closed extremal orbits on the Fermi surface, which contribute to the QO signals, can be visualized. When the magnetic field is along the $c$-axis,  the warped cylindrical Fermi sheet of Band~67 around A--$\Gamma$--A contributes two extremal orbits: a neck orbit ($\gamma$) at the zone centre $k_z=0$, and a belly orbit ($\delta$) around the A point located at the zone boundary $k_z=0.5$ (see orange and yellow orbits in Fig.~\ref{fig4}(c)), respectively. Meanwhile, the large hexagon-like orbit around the A point corresponds to $\zeta$ (red). The largest set of three nearly circular orbits at $k_z=0$ and $k_z=\pm0.08$, experimentally undetected, corresponds to $\sim$11,500~T (green and blue). Finally, for Band~67, there are six small elliptical orbits at $k_z=0$ with a calculated frequency of 103~T (purple), a set of three crescents at $k_z=0.5$ around the H point (light blue), and similar sets of three crescents at $k_z=\pm0.41$ (pink). The calculated frequencies for these crescents at $k_z=0.5$ and $k_z=\pm0.41$ are 96~T and 82~T (Table~\ref{table1}). With regard to Band~68, small oval orbits at $k_z=0.5$ with 112~T (violet) are revealed in the calculation (Fig.~\ref{fig4}(d)). For clarity, only the orbits with positive $k_z$ are shown, and for the crescents, only one set each at $k_z=0.5$ and $k_z=+0.41$ is displayed in Fig.~\ref{fig4}(c). Experimentally, $\alpha$ and $\beta$ are the smallest frequencies detected. To assign $\alpha$ and $\beta$ definitively to the small frequencies from the calculations is challenging, and the possibility that $\beta$ could be a harmonic of $\alpha$ introduces further complications (see Table~S2). Nevertheless, the resolved $\alpha$ and $\beta$ can be good representations of these small extremal orbits, but a larger field range is needed to fully distinguish these small, closely-spaced frequencies in the future. 
\\\\

\noindent{\bf \large{Discussion}}

\noindent After establishing the correspondence between the detected QO frequencies and the calculated results, we further compare the effective masses. The experimentally determined quasiparticle effective masses at both 27.0~kbar and 33.8~kbar are all larger than the calculated band masses at 30~kbar, as tabulated in Table~\ref{table1}. Since both 27.0~kbar and 33.8~kbar are still close to the border of the CDW phase, the larger effective masses obtained by our QO measurements could be attributed to enhanced quantum fluctuations, as discussed earlier. We note that the value of $(m^*/m^*_{cal})$ for all observed frequencies near 30~kbar is less than 2 (Table~\ref{table1}). Such a modest ratio could originate from electron-phonon coupling, which may play a role in mediating superconductivity in CsV$_3$Sb$_5$. Indeed, electron-phonon coupling as the superconducting pairing mechanism in CsV$_3$Sb$_5$ has been explored by other groups~\cite{Zhong2023,Wang2023c}. To further explore if CDW fluctuations are the dominant factor, quantum oscillations at higher pressures as well as definitive DFT results in the CDW phase will be useful, so that $(m^*/m^*_{cal})$ can be tracked over a large pressure range covering both the pristine and the CDW phases.

Despite the modest $(m^*/m^*_{cal})$, the ratio appears to appreciate on approaching $p_c\sim20~$kbar, where $T_{\rm CDW}\rightarrow 0$, indicating that the phenomenon can be tied to the destabilization of the CDW phase and thus the fluctuations associated with the ordered phase. The fact that $T_c$ is also maximized near $p_c$, {\it i.e.} the presence of the second dome around $p_c$, suggests that superconducting pairings can take advantage of enhanced fluctuations. This scenario is in stark contrast to cases in which $T_c$ does not show a dome-like pressure dependence when the CDW phase is suppressed, see for instance the $T$-$p$ phase diagram of 1T-TaS$_2$~\cite{Sipos2008}.

The warped cylinder of Band~67 is known to be robust against the formation of the CDW order~\cite{Tsirlin2022}. Indeed, $\delta$ and $\gamma$, the belly and the neck of the warped cylinder, are detected over the entire pressure range studied. The smooth variation of both frequencies enables us to discuss the pressure-evolution of the shape of this cylinder. With increasing pressure, $\delta$ increases while $\gamma$ decreases, indicating that this Fermi surface sheet becomes more 3D-like under pressure. The increased three-dimensionality under pressure implies that the hopping integral perpendicular to the kagome layers increases, and the pressure evolution of this particular cylinder is consistent with recent theoretical studies~\cite{Tsirlin2022,Wang2022,Wenzel2023}, which predict that $\gamma$ will continue to shrink and eventually disappears at a higher pressure. The pressure evolution of the warped cylinder is important for understanding both the CDW and the superconducting properties. For instance, DFT calculations have linked the disappearance of $\gamma$ to the complete suppression of superconductivity at around 120~kbar~\cite{Tsirlin2022}. Thus, our data provide the experimental foundation for existing DFT results that explore the pressure-dependent electronic structure of \CsVSb.
\\

\noindent{\bf \large{Materials and Methods}}

\noindent{{\bf Crystal growths}} High-quality single crystals of \CsVSb~were synthesized from Cs (ingot, 99.95 $\%$), V (powder, 99.9 $\%$) and Sb (shot, 99.9999 $\%$) using self-flux method~\cite{Ortiz2019,Ortiz2020}. The raw materials were sealed inside a pure-Ar-filled stainless steel jacket with the molar ratio of Cs:V:Sb = 7:3:14. The as-grown single crystals were millimeter-sized
shiny plates.\\

\noindent{{\bf High-pressure quantum oscillation measurements}} Magnetotransport properties were measured by a standard four-terminal configuration in a Physical Property Measurement System by Quantum Design. The high field transport properties were measured at LNCMI
Grenoble using a $^3$He cryostat up to 29 T. The high-pressure environment was achieved by diamond anvil cell (DAC) with patterned electrodes on top of the anvil. A thin flake exfoliated from the bulk \CsVSb\ was transferred onto the patterned electrodes. A layer of h-BN was added above the thin flake for encapsulation and the thickness of the thin flakes was determined by a dual-beam focused ion beam system (Scios~2 DualBeam by Thermo Scientific)~\cite{Xie2021,Ku2022}. The pressure achieved was determined with ruby fluorescence method at room temperature (Fig.~S4). A Stanford Research 830 lock-in amplifier was used for QO measurements.
\\

\noindent{{\bf Band structure calculations}}
The WIEN2k package~\cite{schwarz2003solid}, which employing DFT with full-electron full-potential linearized augmented plane waves, was used in the Fermi surface calculation. The experimental lattice constants for pressurised \CsVSb~\cite{Tsirlin2022} were adopted and no structural optimization was performed.  The generalized gradient approximation of Perdew, Burke and Ernzerhof~\cite{perdew1996generalized} for the exchange-correlation potential, $R_{MT}^{\min}K_{\max}=7.5$, $k$-point mesh of 20,000 in the first Brillouin zone were utilized in the calculation. The muffin-tin radius was set to 2.5 a.u. for Cs and Sb atoms, and 2.47 a.u. for V atoms. Quantum oscillation frequencies were extracted from as-calculated band structures without any energy shift using the Supercell K-space Extremal Area Finder (SKEAF)~\cite{julian2012numerical}, where the averaging threshold for the fractional difference in frequencies and the orbital distances was set to 0.01 and 0.05 respectively.
\\

\noindent{\bf \large{Acknowledgements}}

\noindent{We acknowledge Yoshinori Haga and Chongze Wang for discussions, and Alexander Tsirlin for providing structural parameters under pressure. The work was supported by Research Grants Council of Hong Kong (A-CUHK 402/19, CUHK 14301020, CUHK 14300722), CUHK Direct Grant (4053577, 4053525), City University of Hong Kong (9610438), French National Agency for Research (ANR) within the project FETTOM (ANR-19-CE30-0037), the National Natural Science Foundation of China (Grants No. 12104384 and No.
12174175), and Guangdong Basic and Applied Basic Research Foundation (2022B1515120014). We acknowledge the support of the LNCMI-CNRS, a member of the European Magnetic Field Laboratory (EMFL).
\\

\noindent{\bf \large{Author contributions}}

\noindent{S.K.G. proposed and supervised the project. 
W.Z., T.F.P., A.P., G.S. and G.K. conducted measurements. 
C.W.T., S.T.L. and K.T.L. prepared single crystals of \CsVSb. 
W.W., X.L. and J.X. provided assistance with pressure cells. 
S.W. provided h-BN. 
W.Z., T.F.P. and S.K.G. performed data analysis. W.C.Y. performed the DFT calculations.
W.Z., T.F.P. and S.K.G. wrote the manuscript with input from all authors. }\\

\noindent{\bf \large{Competing interests}}

\noindent{The authors declare no competing interests.}\\

\noindent{\bf \large{Data availability}}

\noindent All study data are included in the article and/or
supporting information.

\newpage

\providecommand{\noopsort}[1]{}\providecommand{\singleletter}[1]{#1}%


\begin{thebibliography}{58}%
\makeatletter
\providecommand \@ifxundefined [1]{%
 \@ifx{#1\undefined}
}%
\providecommand \@ifnum [1]{%
 \ifnum #1\expandafter \@firstoftwo
 \else \expandafter \@secondoftwo
 \fi
}%
\providecommand \@ifx [1]{%
 \ifx #1\expandafter \@firstoftwo
 \else \expandafter \@secondoftwo
 \fi
}%
\providecommand \natexlab [1]{#1}%
\providecommand \enquote  [1]{``#1''}%
\providecommand \bibnamefont  [1]{#1}%
\providecommand \bibfnamefont [1]{#1}%
\providecommand \citenamefont [1]{#1}%
\providecommand \href@noop [0]{\@secondoftwo}%
\providecommand \href [0]{\begingroup \@sanitize@url \@href}%
\providecommand \@href[1]{\@@startlink{#1}\@@href}%
\providecommand \@@href[1]{\endgroup#1\@@endlink}%
\providecommand \@sanitize@url [0]{\catcode `\\12\catcode `\$12\catcode `\&12\catcode `\#12\catcode `\^12\catcode `\_12\catcode `\%12\relax}%
\providecommand \@@startlink[1]{}%
\providecommand \@@endlink[0]{}%
\providecommand \url  [0]{\begingroup\@sanitize@url \@url }%
\providecommand \@url [1]{\endgroup\@href {#1}{\urlprefix }}%
\providecommand \urlprefix  [0]{URL }%
\providecommand \Eprint [0]{\href }%
\providecommand \doibase [0]{http://dx.doi.org/}%
\providecommand \selectlanguage [0]{\@gobble}%
\providecommand \bibinfo  [0]{\@secondoftwo}%
\providecommand \bibfield  [0]{\@secondoftwo}%
\providecommand \translation [1]{[#1]}%
\providecommand \BibitemOpen [0]{}%
\providecommand \bibitemStop [0]{}%
\providecommand \bibitemNoStop [0]{.\EOS\space}%
\providecommand \EOS [0]{\spacefactor3000\relax}%
\providecommand \BibitemShut  [1]{\csname bibitem#1\endcsname}%
\let\auto@bib@innerbib\@empty
\bibitem [{\citenamefont {Ortiz}\ \emph {et~al.}(2019)\citenamefont {Ortiz}, \citenamefont {Gomes}, \citenamefont {Morey}, \citenamefont {Winiarski}, \citenamefont {Bordelon}, \citenamefont {Mangum}, \citenamefont {Oswald}, \citenamefont {Rodriguez-Rivera}, \citenamefont {Neilson}, \citenamefont {Wilson}, \citenamefont {Ertekin}, \citenamefont {McQueen},\ and\ \citenamefont {Toberer}}]{Ortiz2019}%
  \BibitemOpen
  \bibfield  {author} {\bibinfo {author} {\bibfnamefont {B.~R.}\ \bibnamefont {Ortiz}}, \bibinfo {author} {\bibfnamefont {L.~C.}\ \bibnamefont {Gomes}}, \bibinfo {author} {\bibfnamefont {J.~R.}\ \bibnamefont {Morey}}, \bibinfo {author} {\bibfnamefont {M.}~\bibnamefont {Winiarski}}, \bibinfo {author} {\bibfnamefont {M.}~\bibnamefont {Bordelon}}, \bibinfo {author} {\bibfnamefont {J.~S.}\ \bibnamefont {Mangum}}, \bibinfo {author} {\bibfnamefont {I.~W.~H.}\ \bibnamefont {Oswald}}, \bibinfo {author} {\bibfnamefont {J.~A.}\ \bibnamefont {Rodriguez-Rivera}}, \bibinfo {author} {\bibfnamefont {J.~R.}\ \bibnamefont {Neilson}}, \bibinfo {author} {\bibfnamefont {S.~D.}\ \bibnamefont {Wilson}}, \bibinfo {author} {\bibfnamefont {E.}~\bibnamefont {Ertekin}}, \bibinfo {author} {\bibfnamefont {T.~M.}\ \bibnamefont {McQueen}}, \ and\ \bibinfo {author} {\bibfnamefont {E.~S.}\ \bibnamefont {Toberer}},\ }\bibfield  {title} {\enquote {\bibinfo {title} {New kagome prototype materials: discovery of {KV$_3$Sb$_5$}, {RbV$_3$Sb$_5$},
  and {CsV$_3$Sb$_5$}},}\ }\href {\doibase 10.1103/PhysRevMaterials.3.094407} {\bibfield  {journal} {\bibinfo  {journal} {Phys. Rev. Mater.}\ }\textbf {\bibinfo {volume} {3}},\ \bibinfo {pages} {094407} (\bibinfo {year} {2019})}\BibitemShut {NoStop}%
\bibitem [{\citenamefont {Ortiz}\ \emph {et~al.}(2020)\citenamefont {Ortiz}, \citenamefont {Teicher}, \citenamefont {Hu}, \citenamefont {Zuo}, \citenamefont {Sarte}, \citenamefont {Schueller}, \citenamefont {Abeykoon}, \citenamefont {Krogstad}, \citenamefont {Rosenkranz}, \citenamefont {Osborn}, \citenamefont {Seshadri}, \citenamefont {Balents}, \citenamefont {He},\ and\ \citenamefont {Wilson}}]{Ortiz2020}%
  \BibitemOpen
  \bibfield  {author} {\bibinfo {author} {\bibfnamefont {B.~R.}\ \bibnamefont {Ortiz}}, \bibinfo {author} {\bibfnamefont {S.~M.~L.}\ \bibnamefont {Teicher}}, \bibinfo {author} {\bibfnamefont {Y.}~\bibnamefont {Hu}}, \bibinfo {author} {\bibfnamefont {J.~L.}\ \bibnamefont {Zuo}}, \bibinfo {author} {\bibfnamefont {P.~M.}\ \bibnamefont {Sarte}}, \bibinfo {author} {\bibfnamefont {E.~C.}\ \bibnamefont {Schueller}}, \bibinfo {author} {\bibfnamefont {A.~M.~M.}\ \bibnamefont {Abeykoon}}, \bibinfo {author} {\bibfnamefont {M.~J.}\ \bibnamefont {Krogstad}}, \bibinfo {author} {\bibfnamefont {S.}~\bibnamefont {Rosenkranz}}, \bibinfo {author} {\bibfnamefont {R.}~\bibnamefont {Osborn}}, \bibinfo {author} {\bibfnamefont {R.}~\bibnamefont {Seshadri}}, \bibinfo {author} {\bibfnamefont {L.}~\bibnamefont {Balents}}, \bibinfo {author} {\bibfnamefont {J.}~\bibnamefont {He}}, \ and\ \bibinfo {author} {\bibfnamefont {S.~D.}\ \bibnamefont {Wilson}},\ }\bibfield  {title} {\enquote {\bibinfo {title} {{CsV$_3$Sb$_5$}: A
  {${\mathbb{Z}}_{2}$} topological kagome metal with a superconducting ground state},}\ }\href {\doibase 10.1103/PhysRevLett.125.247002} {\bibfield  {journal} {\bibinfo  {journal} {Phys. Rev. Lett.}\ }\textbf {\bibinfo {volume} {125}},\ \bibinfo {pages} {247002} (\bibinfo {year} {2020})}\BibitemShut {NoStop}%
\bibitem [{\citenamefont {Ortiz}\ \emph {et~al.}(2021)\citenamefont {Ortiz}, \citenamefont {Teicher}, \citenamefont {Kautzsch}, \citenamefont {Sarte}, \citenamefont {Ratcliff}, \citenamefont {Harter}, \citenamefont {Ruff}, \citenamefont {Seshadri},\ and\ \citenamefont {Wilson}}]{Ortiz2021}%
  \BibitemOpen
  \bibfield  {author} {\bibinfo {author} {\bibfnamefont {B.~R.}\ \bibnamefont {Ortiz}}, \bibinfo {author} {\bibfnamefont {S.~M.}\ \bibnamefont {Teicher}}, \bibinfo {author} {\bibfnamefont {L.}~\bibnamefont {Kautzsch}}, \bibinfo {author} {\bibfnamefont {P.~M.}\ \bibnamefont {Sarte}}, \bibinfo {author} {\bibfnamefont {N.}~\bibnamefont {Ratcliff}}, \bibinfo {author} {\bibfnamefont {J.}~\bibnamefont {Harter}}, \bibinfo {author} {\bibfnamefont {J.~P.}\ \bibnamefont {Ruff}}, \bibinfo {author} {\bibfnamefont {R.}~\bibnamefont {Seshadri}}, \ and\ \bibinfo {author} {\bibfnamefont {S.~D.}\ \bibnamefont {Wilson}},\ }\bibfield  {title} {\enquote {\bibinfo {title} {Fermi surface mapping and the nature of charge-density-wave order in the kagome superconductor {CsV$_3$Sb$_5$}},}\ }\href@noop {} {\bibfield  {journal} {\bibinfo  {journal} {Phys. Rev. X}\ }\textbf {\bibinfo {volume} {11}},\ \bibinfo {pages} {041030} (\bibinfo {year} {2021})}\BibitemShut {NoStop}%
\bibitem [{\citenamefont {Kiesel}\ and\ \citenamefont {Thomale}(2012)}]{Kiesel2012}%
  \BibitemOpen
  \bibfield  {author} {\bibinfo {author} {\bibfnamefont {M.~L.}\ \bibnamefont {Kiesel}}\ and\ \bibinfo {author} {\bibfnamefont {R.}~\bibnamefont {Thomale}},\ }\bibfield  {title} {\enquote {\bibinfo {title} {Sublattice interference in the kagome {Hubbard} model},}\ }\href {\doibase 10.1103/PhysRevB.86.121105} {\bibfield  {journal} {\bibinfo  {journal} {Phys. Rev. B}\ }\textbf {\bibinfo {volume} {86}},\ \bibinfo {pages} {121105} (\bibinfo {year} {2012})}\BibitemShut {NoStop}%
\bibitem [{\citenamefont {Kiesel}\ \emph {et~al.}(2013)\citenamefont {Kiesel}, \citenamefont {Platt},\ and\ \citenamefont {Thomale}}]{Kiesel2013}%
  \BibitemOpen
  \bibfield  {author} {\bibinfo {author} {\bibfnamefont {M.~L.}\ \bibnamefont {Kiesel}}, \bibinfo {author} {\bibfnamefont {C.}~\bibnamefont {Platt}}, \ and\ \bibinfo {author} {\bibfnamefont {R.}~\bibnamefont {Thomale}},\ }\bibfield  {title} {\enquote {\bibinfo {title} {Unconventional {Fermi} surface instabilities in the kagome {Hubbard} model},}\ }\href {\doibase 10.1103/PhysRevLett.110.126405} {\bibfield  {journal} {\bibinfo  {journal} {Phys. Rev. Lett.}\ }\textbf {\bibinfo {volume} {110}},\ \bibinfo {pages} {126405} (\bibinfo {year} {2013})}\BibitemShut {NoStop}%
\bibitem [{\citenamefont {Wang}\ \emph {et~al.}(2013)\citenamefont {Wang}, \citenamefont {Li}, \citenamefont {Xiang},\ and\ \citenamefont {Wang}}]{Wang2013}%
  \BibitemOpen
  \bibfield  {author} {\bibinfo {author} {\bibfnamefont {W.-S.}\ \bibnamefont {Wang}}, \bibinfo {author} {\bibfnamefont {Z.-Z.}\ \bibnamefont {Li}}, \bibinfo {author} {\bibfnamefont {Y.-Y.}\ \bibnamefont {Xiang}}, \ and\ \bibinfo {author} {\bibfnamefont {Q.-H.}\ \bibnamefont {Wang}},\ }\bibfield  {title} {\enquote {\bibinfo {title} {Competing electronic orders on kagome lattices at van {Hove} filling},}\ }\href {\doibase 10.1103/PhysRevB.87.115135} {\bibfield  {journal} {\bibinfo  {journal} {Phys. Rev. B}\ }\textbf {\bibinfo {volume} {87}},\ \bibinfo {pages} {115135} (\bibinfo {year} {2013})}\BibitemShut {NoStop}%
\bibitem [{\citenamefont {Nie}\ \emph {et~al.}(2022)\citenamefont {Nie}, \citenamefont {Sun}, \citenamefont {Ma}, \citenamefont {Song}, \citenamefont {Zheng}, \citenamefont {Liang}, \citenamefont {Wu}, \citenamefont {Yu}, \citenamefont {Li}, \citenamefont {Shan}, \citenamefont {Zhao}, \citenamefont {Li}, \citenamefont {Kang}, \citenamefont {Wu}, \citenamefont {Zhou}, \citenamefont {Liu}, \citenamefont {Xiang}, \citenamefont {Ying}, \citenamefont {Wang}, \citenamefont {Wu},\ and\ \citenamefont {Chen}}]{Nie2022}%
  \BibitemOpen
  \bibfield  {author} {\bibinfo {author} {\bibfnamefont {L.}~\bibnamefont {Nie}}, \bibinfo {author} {\bibfnamefont {K.}~\bibnamefont {Sun}}, \bibinfo {author} {\bibfnamefont {W.}~\bibnamefont {Ma}}, \bibinfo {author} {\bibfnamefont {D.}~\bibnamefont {Song}}, \bibinfo {author} {\bibfnamefont {L.}~\bibnamefont {Zheng}}, \bibinfo {author} {\bibfnamefont {Z.}~\bibnamefont {Liang}}, \bibinfo {author} {\bibfnamefont {P.}~\bibnamefont {Wu}}, \bibinfo {author} {\bibfnamefont {F.}~\bibnamefont {Yu}}, \bibinfo {author} {\bibfnamefont {J.}~\bibnamefont {Li}}, \bibinfo {author} {\bibfnamefont {M.}~\bibnamefont {Shan}}, \bibinfo {author} {\bibfnamefont {D.}~\bibnamefont {Zhao}}, \bibinfo {author} {\bibfnamefont {S.}~\bibnamefont {Li}}, \bibinfo {author} {\bibfnamefont {B.}~\bibnamefont {Kang}}, \bibinfo {author} {\bibfnamefont {Z.}~\bibnamefont {Wu}}, \bibinfo {author} {\bibfnamefont {Y.}~\bibnamefont {Zhou}}, \bibinfo {author} {\bibfnamefont {K.}~\bibnamefont {Liu}}, \bibinfo {author} {\bibfnamefont {Z.}~\bibnamefont
  {Xiang}}, \bibinfo {author} {\bibfnamefont {J.}~\bibnamefont {Ying}}, \bibinfo {author} {\bibfnamefont {Z.}~\bibnamefont {Wang}}, \bibinfo {author} {\bibfnamefont {T.}~\bibnamefont {Wu}}, \ and\ \bibinfo {author} {\bibfnamefont {X.}~\bibnamefont {Chen}},\ }\bibfield  {title} {\enquote {\bibinfo {title} {Charge-density-wave-driven electronic nematicity in a kagome superconductor},}\ }\href@noop {} {\bibfield  {journal} {\bibinfo  {journal} {Nature}\ }\textbf {\bibinfo {volume} {604}},\ \bibinfo {pages} {59} (\bibinfo {year} {2022})}\BibitemShut {NoStop}%
\bibitem [{\citenamefont {Li}\ \emph {et~al.}(2022)\citenamefont {Li}, \citenamefont {Zhao}, \citenamefont {Ortiz}, \citenamefont {Park}, \citenamefont {Ye}, \citenamefont {Balents}, \citenamefont {Wang}, \citenamefont {Wilson},\ and\ \citenamefont {Zeljkovic}}]{Li2022}%
  \BibitemOpen
  \bibfield  {author} {\bibinfo {author} {\bibfnamefont {H.}~\bibnamefont {Li}}, \bibinfo {author} {\bibfnamefont {H.}~\bibnamefont {Zhao}}, \bibinfo {author} {\bibfnamefont {B.~R.}\ \bibnamefont {Ortiz}}, \bibinfo {author} {\bibfnamefont {T.}~\bibnamefont {Park}}, \bibinfo {author} {\bibfnamefont {M.}~\bibnamefont {Ye}}, \bibinfo {author} {\bibfnamefont {L.}~\bibnamefont {Balents}}, \bibinfo {author} {\bibfnamefont {Z.}~\bibnamefont {Wang}}, \bibinfo {author} {\bibfnamefont {S.~D.}\ \bibnamefont {Wilson}}, \ and\ \bibinfo {author} {\bibfnamefont {I.}~\bibnamefont {Zeljkovic}},\ }\bibfield  {title} {\enquote {\bibinfo {title} {Rotation symmetry breaking in the normal state of a kagome superconductor {KV$_3$Sb$_5$}},}\ }\href@noop {} {\bibfield  {journal} {\bibinfo  {journal} {Nat. Phys.}\ }\textbf {\bibinfo {volume} {18}},\ \bibinfo {pages} {265} (\bibinfo {year} {2022})}\BibitemShut {NoStop}%
\bibitem [{\citenamefont {Xu}\ \emph {et~al.}(2022)\citenamefont {Xu}, \citenamefont {Ni}, \citenamefont {Liu}, \citenamefont {Ortiz}, \citenamefont {Deng}, \citenamefont {Wilson}, \citenamefont {Yan}, \citenamefont {Balents},\ and\ \citenamefont {Wu}}]{Xu2022}%
  \BibitemOpen
  \bibfield  {author} {\bibinfo {author} {\bibfnamefont {Y.}~\bibnamefont {Xu}}, \bibinfo {author} {\bibfnamefont {Z.}~\bibnamefont {Ni}}, \bibinfo {author} {\bibfnamefont {Y.}~\bibnamefont {Liu}}, \bibinfo {author} {\bibfnamefont {B.~R.}\ \bibnamefont {Ortiz}}, \bibinfo {author} {\bibfnamefont {Q.}~\bibnamefont {Deng}}, \bibinfo {author} {\bibfnamefont {S.~D.}\ \bibnamefont {Wilson}}, \bibinfo {author} {\bibfnamefont {B.}~\bibnamefont {Yan}}, \bibinfo {author} {\bibfnamefont {L.}~\bibnamefont {Balents}}, \ and\ \bibinfo {author} {\bibfnamefont {L.}~\bibnamefont {Wu}},\ }\bibfield  {title} {\enquote {\bibinfo {title} {Three-state nematicity and magneto-optical {Kerr} effect in the charge density waves in kagome superconductors},}\ }\href@noop {} {\bibfield  {journal} {\bibinfo  {journal} {Nat. Phys.}\ }\textbf {\bibinfo {volume} {18}},\ \bibinfo {pages} {1470} (\bibinfo {year} {2022})}\BibitemShut {NoStop}%
\bibitem [{\citenamefont {Yin}\ \emph {et~al.}(2021)\citenamefont {Yin}, \citenamefont {Tu}, \citenamefont {Gong}, \citenamefont {Fu}, \citenamefont {Yan},\ and\ \citenamefont {Lei}}]{Yin2021}%
  \BibitemOpen
  \bibfield  {author} {\bibinfo {author} {\bibfnamefont {Q.}~\bibnamefont {Yin}}, \bibinfo {author} {\bibfnamefont {Z.}~\bibnamefont {Tu}}, \bibinfo {author} {\bibfnamefont {C.}~\bibnamefont {Gong}}, \bibinfo {author} {\bibfnamefont {Y.}~\bibnamefont {Fu}}, \bibinfo {author} {\bibfnamefont {S.}~\bibnamefont {Yan}}, \ and\ \bibinfo {author} {\bibfnamefont {H.}~\bibnamefont {Lei}},\ }\bibfield  {title} {\enquote {\bibinfo {title} {Superconductivity and normal-state properties of kagome metal {RbV$_3$Sb$_5$} single crystals},}\ }\href@noop {} {\bibfield  {journal} {\bibinfo  {journal} {Chin. Phys. Lett.}\ }\textbf {\bibinfo {volume} {38}},\ \bibinfo {pages} {037403} (\bibinfo {year} {2021})}\BibitemShut {NoStop}%
\bibitem [{\citenamefont {Tan}\ \emph {et~al.}(2021)\citenamefont {Tan}, \citenamefont {Liu}, \citenamefont {Wang},\ and\ \citenamefont {Yan}}]{Tan2021}%
  \BibitemOpen
  \bibfield  {author} {\bibinfo {author} {\bibfnamefont {H.}~\bibnamefont {Tan}}, \bibinfo {author} {\bibfnamefont {Y.}~\bibnamefont {Liu}}, \bibinfo {author} {\bibfnamefont {Z.}~\bibnamefont {Wang}}, \ and\ \bibinfo {author} {\bibfnamefont {B.}~\bibnamefont {Yan}},\ }\bibfield  {title} {\enquote {\bibinfo {title} {Charge density waves and electronic properties of superconducting kagome metals},}\ }\href@noop {} {\bibfield  {journal} {\bibinfo  {journal} {Phys. Rev. Lett.}\ }\textbf {\bibinfo {volume} {127}},\ \bibinfo {pages} {046401} (\bibinfo {year} {2021})}\BibitemShut {NoStop}%
\bibitem [{\citenamefont {Wang}\ \emph {et~al.}(2023{\natexlab{a}})\citenamefont {Wang}, \citenamefont {Zhang}, \citenamefont {Wang}, \citenamefont {Poon}, \citenamefont {Wang}, \citenamefont {Tsang}, \citenamefont {Xie}, \citenamefont {Zhou}, \citenamefont {Zhao}, \citenamefont {Wang} \emph {et~al.}}]{Wang2023}%
  \BibitemOpen
  \bibfield  {author} {\bibinfo {author} {\bibfnamefont {L.}~\bibnamefont {Wang}}, \bibinfo {author} {\bibfnamefont {W.}~\bibnamefont {Zhang}}, \bibinfo {author} {\bibfnamefont {Z.}~\bibnamefont {Wang}}, \bibinfo {author} {\bibfnamefont {T.~F.}\ \bibnamefont {Poon}}, \bibinfo {author} {\bibfnamefont {W.}~\bibnamefont {Wang}}, \bibinfo {author} {\bibfnamefont {C.~W.}\ \bibnamefont {Tsang}}, \bibinfo {author} {\bibfnamefont {J.}~\bibnamefont {Xie}}, \bibinfo {author} {\bibfnamefont {X.}~\bibnamefont {Zhou}}, \bibinfo {author} {\bibfnamefont {Y.}~\bibnamefont {Zhao}}, \bibinfo {author} {\bibfnamefont {S.}~\bibnamefont {Wang}},  \emph {et~al.},\ }\bibfield  {title} {\enquote {\bibinfo {title} {Anomalous {Hall} effect and two-dimensional {Fermi} surfaces in the charge-density-wave state of kagome metal {RbV$_3$Sb$_5$}},}\ }\href@noop {} {\bibfield  {journal} {\bibinfo  {journal} {J. Phys. Mater.}\ }\textbf {\bibinfo {volume} {6}},\ \bibinfo {pages} {02LT01} (\bibinfo {year} {2023}{\natexlab{a}})}\BibitemShut
  {NoStop}%
\bibitem [{\citenamefont {Wang}\ \emph {et~al.}(2023{\natexlab{b}})\citenamefont {Wang}, \citenamefont {Zhang}, \citenamefont {Wang}, \citenamefont {Poon}, \citenamefont {Tsang}, \citenamefont {Wang}, \citenamefont {Xie}, \citenamefont {Lam}, \citenamefont {Zhou}, \citenamefont {Zhao}, \citenamefont {Wang}, \citenamefont {Ai}, \citenamefont {Lai},\ and\ \citenamefont {Goh}}]{Wang2023b}%
  \BibitemOpen
  \bibfield  {author} {\bibinfo {author} {\bibfnamefont {Z.}~\bibnamefont {Wang}}, \bibinfo {author} {\bibfnamefont {W.}~\bibnamefont {Zhang}}, \bibinfo {author} {\bibfnamefont {L.}~\bibnamefont {Wang}}, \bibinfo {author} {\bibfnamefont {T.~F.}\ \bibnamefont {Poon}}, \bibinfo {author} {\bibfnamefont {C.~W.}\ \bibnamefont {Tsang}}, \bibinfo {author} {\bibfnamefont {W.}~\bibnamefont {Wang}}, \bibinfo {author} {\bibfnamefont {J.}~\bibnamefont {Xie}}, \bibinfo {author} {\bibfnamefont {S.~T.}\ \bibnamefont {Lam}}, \bibinfo {author} {\bibfnamefont {X.}~\bibnamefont {Zhou}}, \bibinfo {author} {\bibfnamefont {Y.}~\bibnamefont {Zhao}}, \bibinfo {author} {\bibfnamefont {S.}~\bibnamefont {Wang}}, \bibinfo {author} {\bibfnamefont {M.-Z.}\ \bibnamefont {Ai}}, \bibinfo {author} {\bibfnamefont {K.~T.}\ \bibnamefont {Lai}}, \ and\ \bibinfo {author} {\bibfnamefont {S.~K.}\ \bibnamefont {Goh}},\ }\bibfield  {title} {\enquote {\bibinfo {title} {{Similarities and differences in the fermiology of {kagome} metals {AV$_3$Sb$_5$}
  ({A$=$K, Rb, Cs}) revealed by {Shubnikov–de Haas} oscillations}},}\ }\href {\doibase 10.1063/5.0145859} {\bibfield  {journal} {\bibinfo  {journal} {Appl. Phys. Lett.}\ }\textbf {\bibinfo {volume} {123}},\ \bibinfo {pages} {012601} (\bibinfo {year} {2023}{\natexlab{b}})}\BibitemShut {NoStop}%
\bibitem [{\citenamefont {Yang}\ \emph {et~al.}(2020)\citenamefont {Yang}, \citenamefont {Wang}, \citenamefont {Ortiz}, \citenamefont {Liu}, \citenamefont {Gayles}, \citenamefont {Derunova}, \citenamefont {Gonzalez-Hernandez}, \citenamefont {{\v{S}}mejkal}, \citenamefont {Chen}, \citenamefont {Parkin}, \citenamefont {Wilson}, \citenamefont {Toberer}, \citenamefont {McQueen},\ and\ \citenamefont {Ali}}]{Yang2020}%
  \BibitemOpen
  \bibfield  {author} {\bibinfo {author} {\bibfnamefont {S.-Y.}\ \bibnamefont {Yang}}, \bibinfo {author} {\bibfnamefont {Y.}~\bibnamefont {Wang}}, \bibinfo {author} {\bibfnamefont {B.~R.}\ \bibnamefont {Ortiz}}, \bibinfo {author} {\bibfnamefont {D.}~\bibnamefont {Liu}}, \bibinfo {author} {\bibfnamefont {J.}~\bibnamefont {Gayles}}, \bibinfo {author} {\bibfnamefont {E.}~\bibnamefont {Derunova}}, \bibinfo {author} {\bibfnamefont {R.}~\bibnamefont {Gonzalez-Hernandez}}, \bibinfo {author} {\bibfnamefont {L.}~\bibnamefont {{\v{S}}mejkal}}, \bibinfo {author} {\bibfnamefont {Y.}~\bibnamefont {Chen}}, \bibinfo {author} {\bibfnamefont {S.~S.~P.}\ \bibnamefont {Parkin}}, \bibinfo {author} {\bibfnamefont {S.~D.}\ \bibnamefont {Wilson}}, \bibinfo {author} {\bibfnamefont {E.~S.}\ \bibnamefont {Toberer}}, \bibinfo {author} {\bibfnamefont {T.}~\bibnamefont {McQueen}}, \ and\ \bibinfo {author} {\bibfnamefont {M.~N.}\ \bibnamefont {Ali}},\ }\bibfield  {title} {\enquote {\bibinfo {title} {Giant, unconventional anomalous {Hall}
  effect in the metallic frustrated magnet candidate, {KV$_3$Sb$_5$}},}\ }\href@noop {} {\bibfield  {journal} {\bibinfo  {journal} {Sci. Adv.}\ }\textbf {\bibinfo {volume} {6}},\ \bibinfo {pages} {eabb6003} (\bibinfo {year} {2020})}\BibitemShut {NoStop}%
\bibitem [{\citenamefont {Du}\ \emph {et~al.}(2021)\citenamefont {Du}, \citenamefont {Luo}, \citenamefont {Ortiz}, \citenamefont {Chen}, \citenamefont {Duan}, \citenamefont {Zhang}, \citenamefont {Lu}, \citenamefont {Wilson}, \citenamefont {Song},\ and\ \citenamefont {Yuan}}]{Du2021}%
  \BibitemOpen
  \bibfield  {author} {\bibinfo {author} {\bibfnamefont {F.}~\bibnamefont {Du}}, \bibinfo {author} {\bibfnamefont {S.}~\bibnamefont {Luo}}, \bibinfo {author} {\bibfnamefont {B.~R.}\ \bibnamefont {Ortiz}}, \bibinfo {author} {\bibfnamefont {Y.}~\bibnamefont {Chen}}, \bibinfo {author} {\bibfnamefont {W.}~\bibnamefont {Duan}}, \bibinfo {author} {\bibfnamefont {D.}~\bibnamefont {Zhang}}, \bibinfo {author} {\bibfnamefont {X.}~\bibnamefont {Lu}}, \bibinfo {author} {\bibfnamefont {S.~D.}\ \bibnamefont {Wilson}}, \bibinfo {author} {\bibfnamefont {Y.}~\bibnamefont {Song}}, \ and\ \bibinfo {author} {\bibfnamefont {H.}~\bibnamefont {Yuan}},\ }\bibfield  {title} {\enquote {\bibinfo {title} {Pressure-induced double superconducting domes and charge instability in the kagome metal {KV$_3$Sb$_5$}},}\ }\href {\doibase 10.1103/PhysRevB.103.L220504} {\bibfield  {journal} {\bibinfo  {journal} {Phys. Rev. B}\ }\textbf {\bibinfo {volume} {103}},\ \bibinfo {pages} {L220504} (\bibinfo {year} {2021})}\BibitemShut {NoStop}%
\bibitem [{\citenamefont {Wang}\ \emph {et~al.}(2021)\citenamefont {Wang}, \citenamefont {Chen}, \citenamefont {Yin}, \citenamefont {Ma}, \citenamefont {Pan}, \citenamefont {Yang}, \citenamefont {Ji}, \citenamefont {Wu}, \citenamefont {Shan}, \citenamefont {Xu}, \citenamefont {Tu}, \citenamefont {Gong}, \citenamefont {Liu}, \citenamefont {Li}, \citenamefont {Uwatoko}, \citenamefont {Dong}, \citenamefont {Lei}, \citenamefont {Sun},\ and\ \citenamefont {Cheng}}]{Wang2021a}%
  \BibitemOpen
  \bibfield  {author} {\bibinfo {author} {\bibfnamefont {N.~N.}\ \bibnamefont {Wang}}, \bibinfo {author} {\bibfnamefont {K.~Y.}\ \bibnamefont {Chen}}, \bibinfo {author} {\bibfnamefont {Q.~W.}\ \bibnamefont {Yin}}, \bibinfo {author} {\bibfnamefont {Y.~N.~N.}\ \bibnamefont {Ma}}, \bibinfo {author} {\bibfnamefont {B.~Y.}\ \bibnamefont {Pan}}, \bibinfo {author} {\bibfnamefont {X.}~\bibnamefont {Yang}}, \bibinfo {author} {\bibfnamefont {X.~Y.}\ \bibnamefont {Ji}}, \bibinfo {author} {\bibfnamefont {S.~L.}\ \bibnamefont {Wu}}, \bibinfo {author} {\bibfnamefont {P.~F.}\ \bibnamefont {Shan}}, \bibinfo {author} {\bibfnamefont {S.~X.}\ \bibnamefont {Xu}}, \bibinfo {author} {\bibfnamefont {Z.~J.}\ \bibnamefont {Tu}}, \bibinfo {author} {\bibfnamefont {C.~S.}\ \bibnamefont {Gong}}, \bibinfo {author} {\bibfnamefont {G.~T.}\ \bibnamefont {Liu}}, \bibinfo {author} {\bibfnamefont {G.}~\bibnamefont {Li}}, \bibinfo {author} {\bibfnamefont {Y.}~\bibnamefont {Uwatoko}}, \bibinfo {author} {\bibfnamefont {X.~L.}\ \bibnamefont
  {Dong}}, \bibinfo {author} {\bibfnamefont {H.~C.}\ \bibnamefont {Lei}}, \bibinfo {author} {\bibfnamefont {J.~P.}\ \bibnamefont {Sun}}, \ and\ \bibinfo {author} {\bibfnamefont {J.-G.}\ \bibnamefont {Cheng}},\ }\bibfield  {title} {\enquote {\bibinfo {title} {Competition between charge-density-wave and superconductivity in the kagome metal {RbV$_3$Sb$_5$}},}\ }\href {\doibase 10.1103/PhysRevResearch.3.043018} {\bibfield  {journal} {\bibinfo  {journal} {Phys. Rev. Res.}\ }\textbf {\bibinfo {volume} {3}},\ \bibinfo {pages} {043018} (\bibinfo {year} {2021})}\BibitemShut {NoStop}%
\bibitem [{\citenamefont {Zheng}\ \emph {et~al.}(2022)\citenamefont {Zheng}, \citenamefont {Wu}, \citenamefont {Yang}, \citenamefont {Nie}, \citenamefont {Shan}, \citenamefont {Sun}, \citenamefont {Song}, \citenamefont {Yu}, \citenamefont {Li}, \citenamefont {Zhao} \emph {et~al.}}]{Zheng2022}%
  \BibitemOpen
  \bibfield  {author} {\bibinfo {author} {\bibfnamefont {L.}~\bibnamefont {Zheng}}, \bibinfo {author} {\bibfnamefont {Z.}~\bibnamefont {Wu}}, \bibinfo {author} {\bibfnamefont {Y.}~\bibnamefont {Yang}}, \bibinfo {author} {\bibfnamefont {L.}~\bibnamefont {Nie}}, \bibinfo {author} {\bibfnamefont {M.}~\bibnamefont {Shan}}, \bibinfo {author} {\bibfnamefont {K.}~\bibnamefont {Sun}}, \bibinfo {author} {\bibfnamefont {D.}~\bibnamefont {Song}}, \bibinfo {author} {\bibfnamefont {F.}~\bibnamefont {Yu}}, \bibinfo {author} {\bibfnamefont {J.}~\bibnamefont {Li}}, \bibinfo {author} {\bibfnamefont {D.}~\bibnamefont {Zhao}},  \emph {et~al.},\ }\bibfield  {title} {\enquote {\bibinfo {title} {Emergent charge order in pressurized kagome superconductor {CsV$_3$Sb$_5$}},}\ }\href@noop {} {\bibfield  {journal} {\bibinfo  {journal} {Nature}\ }\textbf {\bibinfo {volume} {611}},\ \bibinfo {pages} {682} (\bibinfo {year} {2022})}\BibitemShut {NoStop}%
\bibitem [{\citenamefont {Kang}\ \emph {et~al.}(2022)\citenamefont {Kang}, \citenamefont {Fang}, \citenamefont {Kim}, \citenamefont {Ortiz}, \citenamefont {Ryu}, \citenamefont {Kim}, \citenamefont {Yoo}, \citenamefont {Sangiovanni}, \citenamefont {Di~Sante}, \citenamefont {Park}, \citenamefont {Jozwiak}, \citenamefont {Bostwick}, \citenamefont {Rotenberg}, \citenamefont {Kaxiras}, \citenamefont {Wilson}, \citenamefont {Park},\ and\ \citenamefont {Comin}}]{Kang2022}%
  \BibitemOpen
  \bibfield  {author} {\bibinfo {author} {\bibfnamefont {M.}~\bibnamefont {Kang}}, \bibinfo {author} {\bibfnamefont {S.}~\bibnamefont {Fang}}, \bibinfo {author} {\bibfnamefont {J.-K.}\ \bibnamefont {Kim}}, \bibinfo {author} {\bibfnamefont {B.~R.}\ \bibnamefont {Ortiz}}, \bibinfo {author} {\bibfnamefont {S.~H.}\ \bibnamefont {Ryu}}, \bibinfo {author} {\bibfnamefont {J.}~\bibnamefont {Kim}}, \bibinfo {author} {\bibfnamefont {J.}~\bibnamefont {Yoo}}, \bibinfo {author} {\bibfnamefont {G.}~\bibnamefont {Sangiovanni}}, \bibinfo {author} {\bibfnamefont {D.}~\bibnamefont {Di~Sante}}, \bibinfo {author} {\bibfnamefont {B.-G.}\ \bibnamefont {Park}}, \bibinfo {author} {\bibfnamefont {C.}~\bibnamefont {Jozwiak}}, \bibinfo {author} {\bibfnamefont {A.}~\bibnamefont {Bostwick}}, \bibinfo {author} {\bibfnamefont {E.}~\bibnamefont {Rotenberg}}, \bibinfo {author} {\bibfnamefont {E.}~\bibnamefont {Kaxiras}}, \bibinfo {author} {\bibfnamefont {S.~D.}\ \bibnamefont {Wilson}}, \bibinfo {author} {\bibfnamefont {J.-H.}\ \bibnamefont
  {Park}}, \ and\ \bibinfo {author} {\bibfnamefont {R.}~\bibnamefont {Comin}},\ }\bibfield  {title} {\enquote {\bibinfo {title} {Twofold van {Hove} singularity and origin of charge order in topological kagome superconductor {CsV$_3$Sb$_5$}},}\ }\href@noop {} {\bibfield  {journal} {\bibinfo  {journal} {Nat. Phys.}\ }\textbf {\bibinfo {volume} {18}},\ \bibinfo {pages} {301} (\bibinfo {year} {2022})}\BibitemShut {NoStop}%
\bibitem [{\citenamefont {Feng}\ \emph {et~al.}(2023)\citenamefont {Feng}, \citenamefont {Zhao}, \citenamefont {Luo}, \citenamefont {Yang}, \citenamefont {Fang}, \citenamefont {Yang}, \citenamefont {Gao}, \citenamefont {Zhou},\ and\ \citenamefont {Zheng}}]{Feng2023}%
  \BibitemOpen
  \bibfield  {author} {\bibinfo {author} {\bibfnamefont {X.}~\bibnamefont {Feng}}, \bibinfo {author} {\bibfnamefont {Z.}~\bibnamefont {Zhao}}, \bibinfo {author} {\bibfnamefont {J.}~\bibnamefont {Luo}}, \bibinfo {author} {\bibfnamefont {J.}~\bibnamefont {Yang}}, \bibinfo {author} {\bibfnamefont {A.}~\bibnamefont {Fang}}, \bibinfo {author} {\bibfnamefont {H.}~\bibnamefont {Yang}}, \bibinfo {author} {\bibfnamefont {H.}~\bibnamefont {Gao}}, \bibinfo {author} {\bibfnamefont {R.}~\bibnamefont {Zhou}}, \ and\ \bibinfo {author} {\bibfnamefont {G.-q.}\ \bibnamefont {Zheng}},\ }\bibfield  {title} {\enquote {\bibinfo {title} {Commensurate-to-incommensurate transition of charge-density-wave order and a possible quantum critical point in pressurized kagome metal {CsV$_3$Sb$_5$}},}\ }\href@noop {} {\bibfield  {journal} {\bibinfo  {journal} {npj Quant. Mater.}\ }\textbf {\bibinfo {volume} {8}},\ \bibinfo {pages} {23} (\bibinfo {year} {2023})}\BibitemShut {NoStop}%
\bibitem [{\citenamefont {Asaba}\ \emph {et~al.}(2024)\citenamefont {Asaba}, \citenamefont {Onishi}, \citenamefont {Kageyama}, \citenamefont {Kiyosue}, \citenamefont {Ohtsuka}, \citenamefont {Suetsugu}, \citenamefont {Kohsaka}, \citenamefont {Gaggl}, \citenamefont {Kasahara}, \citenamefont {Murayama} \emph {et~al.}}]{Asaba2023}%
  \BibitemOpen
  \bibfield  {author} {\bibinfo {author} {\bibfnamefont {T.}~\bibnamefont {Asaba}}, \bibinfo {author} {\bibfnamefont {A.}~\bibnamefont {Onishi}}, \bibinfo {author} {\bibfnamefont {Y.}~\bibnamefont {Kageyama}}, \bibinfo {author} {\bibfnamefont {T.}~\bibnamefont {Kiyosue}}, \bibinfo {author} {\bibfnamefont {K.}~\bibnamefont {Ohtsuka}}, \bibinfo {author} {\bibfnamefont {S.}~\bibnamefont {Suetsugu}}, \bibinfo {author} {\bibfnamefont {Y.}~\bibnamefont {Kohsaka}}, \bibinfo {author} {\bibfnamefont {T.}~\bibnamefont {Gaggl}}, \bibinfo {author} {\bibfnamefont {Y.}~\bibnamefont {Kasahara}}, \bibinfo {author} {\bibfnamefont {H.}~\bibnamefont {Murayama}},  \emph {et~al.},\ }\bibfield  {title} {\enquote {\bibinfo {title} {Evidence for an odd-parity nematic phase above the charge density wave transition in kagome metal {CsV$_3$Sb$_5$}},}\ }\href@noop {} {\bibfield  {journal} {\bibinfo  {journal} {Nat. Phys.}\ }\textbf {\bibinfo {volume} {20}},\ \bibinfo {pages} {40} (\bibinfo {year} {2024})}\BibitemShut {NoStop}%
\bibitem [{\citenamefont {Tan}\ \emph {et~al.}(2023)\citenamefont {Tan}, \citenamefont {Li}, \citenamefont {Liu}, \citenamefont {Kaplan}, \citenamefont {Wang},\ and\ \citenamefont {Yan}}]{Tan2023}%
  \BibitemOpen
  \bibfield  {author} {\bibinfo {author} {\bibfnamefont {H.}~\bibnamefont {Tan}}, \bibinfo {author} {\bibfnamefont {Y.}~\bibnamefont {Li}}, \bibinfo {author} {\bibfnamefont {Y.}~\bibnamefont {Liu}}, \bibinfo {author} {\bibfnamefont {D.}~\bibnamefont {Kaplan}}, \bibinfo {author} {\bibfnamefont {Z.}~\bibnamefont {Wang}}, \ and\ \bibinfo {author} {\bibfnamefont {B.}~\bibnamefont {Yan}},\ }\bibfield  {title} {\enquote {\bibinfo {title} {Emergent topological quantum orbits in the charge density wave phase of kagome metal {CsV$_3$Sb$_5$}},}\ }\href@noop {} {\bibfield  {journal} {\bibinfo  {journal} {npj Quant. Mater.}\ }\textbf {\bibinfo {volume} {8}},\ \bibinfo {pages} {39} (\bibinfo {year} {2023})}\BibitemShut {NoStop}%
\bibitem [{\citenamefont {Khasanov}\ \emph {et~al.}(2022)\citenamefont {Khasanov}, \citenamefont {Das}, \citenamefont {Gupta}, \citenamefont {Mielke~III}, \citenamefont {Elender}, \citenamefont {Yin}, \citenamefont {Tu}, \citenamefont {Gong}, \citenamefont {Lei}, \citenamefont {Ritz} \emph {et~al.}}]{Khasanov2022}%
  \BibitemOpen
  \bibfield  {author} {\bibinfo {author} {\bibfnamefont {R.}~\bibnamefont {Khasanov}}, \bibinfo {author} {\bibfnamefont {D.}~\bibnamefont {Das}}, \bibinfo {author} {\bibfnamefont {R.}~\bibnamefont {Gupta}}, \bibinfo {author} {\bibfnamefont {C.}~\bibnamefont {Mielke~III}}, \bibinfo {author} {\bibfnamefont {M.}~\bibnamefont {Elender}}, \bibinfo {author} {\bibfnamefont {Q.}~\bibnamefont {Yin}}, \bibinfo {author} {\bibfnamefont {Z.}~\bibnamefont {Tu}}, \bibinfo {author} {\bibfnamefont {C.}~\bibnamefont {Gong}}, \bibinfo {author} {\bibfnamefont {H.}~\bibnamefont {Lei}}, \bibinfo {author} {\bibfnamefont {E.~T.}\ \bibnamefont {Ritz}},  \emph {et~al.},\ }\bibfield  {title} {\enquote {\bibinfo {title} {Time-reversal symmetry broken by charge order in {CsV$_3$Sb$_5$}},}\ }\href@noop {} {\bibfield  {journal} {\bibinfo  {journal} {Phys. Rev. Res.}\ }\textbf {\bibinfo {volume} {4}},\ \bibinfo {pages} {023244} (\bibinfo {year} {2022})}\BibitemShut {NoStop}%
\bibitem [{\citenamefont {Guo}\ \emph {et~al.}(2022)\citenamefont {Guo}, \citenamefont {Putzke}, \citenamefont {Konyzheva}, \citenamefont {Huang}, \citenamefont {Gutierrez-Amigo}, \citenamefont {Errea}, \citenamefont {Chen}, \citenamefont {Vergniory}, \citenamefont {Felser}, \citenamefont {Fischer} \emph {et~al.}}]{Guo2022}%
  \BibitemOpen
  \bibfield  {author} {\bibinfo {author} {\bibfnamefont {C.}~\bibnamefont {Guo}}, \bibinfo {author} {\bibfnamefont {C.}~\bibnamefont {Putzke}}, \bibinfo {author} {\bibfnamefont {S.}~\bibnamefont {Konyzheva}}, \bibinfo {author} {\bibfnamefont {X.}~\bibnamefont {Huang}}, \bibinfo {author} {\bibfnamefont {M.}~\bibnamefont {Gutierrez-Amigo}}, \bibinfo {author} {\bibfnamefont {I.}~\bibnamefont {Errea}}, \bibinfo {author} {\bibfnamefont {D.}~\bibnamefont {Chen}}, \bibinfo {author} {\bibfnamefont {M.~G.}\ \bibnamefont {Vergniory}}, \bibinfo {author} {\bibfnamefont {C.}~\bibnamefont {Felser}}, \bibinfo {author} {\bibfnamefont {M.~H.}\ \bibnamefont {Fischer}},  \emph {et~al.},\ }\bibfield  {title} {\enquote {\bibinfo {title} {Switchable chiral transport in charge-ordered kagome metal {CsV$_3$Sb$_5$}},}\ }\href@noop {} {\bibfield  {journal} {\bibinfo  {journal} {Nature}\ }\textbf {\bibinfo {volume} {611}},\ \bibinfo {pages} {461} (\bibinfo {year} {2022})}\BibitemShut {NoStop}%
\bibitem [{\citenamefont {Hu}\ \emph {et~al.}(2022{\natexlab{a}})\citenamefont {Hu}, \citenamefont {Yamane}, \citenamefont {Mattoni}, \citenamefont {Yada}, \citenamefont {Obata}, \citenamefont {Li}, \citenamefont {Yao}, \citenamefont {Wang}, \citenamefont {Wang}, \citenamefont {Farhang} \emph {et~al.}}]{Hu2022}%
  \BibitemOpen
  \bibfield  {author} {\bibinfo {author} {\bibfnamefont {Y.}~\bibnamefont {Hu}}, \bibinfo {author} {\bibfnamefont {S.}~\bibnamefont {Yamane}}, \bibinfo {author} {\bibfnamefont {G.}~\bibnamefont {Mattoni}}, \bibinfo {author} {\bibfnamefont {K.}~\bibnamefont {Yada}}, \bibinfo {author} {\bibfnamefont {K.}~\bibnamefont {Obata}}, \bibinfo {author} {\bibfnamefont {Y.}~\bibnamefont {Li}}, \bibinfo {author} {\bibfnamefont {Y.}~\bibnamefont {Yao}}, \bibinfo {author} {\bibfnamefont {Z.}~\bibnamefont {Wang}}, \bibinfo {author} {\bibfnamefont {J.}~\bibnamefont {Wang}}, \bibinfo {author} {\bibfnamefont {C.}~\bibnamefont {Farhang}},  \emph {et~al.},\ }\bibfield  {title} {\enquote {\bibinfo {title} {Time-reversal symmetry breaking in charge density wave of {CsV$_3$Sb$_5$} detected by polar {Kerr} effect},}\ }\href@noop {} {\bibfield  {journal} {\bibinfo  {journal} {arXiv:2208.08036}\ } (\bibinfo {year} {2022}{\natexlab{a}})}\BibitemShut {NoStop}%
\bibitem [{\citenamefont {Yu}\ \emph {et~al.}(2021{\natexlab{a}})\citenamefont {Yu}, \citenamefont {Wang}, \citenamefont {Zhang}, \citenamefont {Sander}, \citenamefont {Ni}, \citenamefont {Lu}, \citenamefont {Ma}, \citenamefont {Wang}, \citenamefont {Zhao}, \citenamefont {Chen}, \citenamefont {Jiang}, \citenamefont {Zhang}, \citenamefont {Yang}, \citenamefont {Zhou}, \citenamefont {Dong}, \citenamefont {Johnson}, \citenamefont {Graf}, \citenamefont {Hu}, \citenamefont {Gao},\ and\ \citenamefont {Zhao}}]{Yu2021c}%
  \BibitemOpen
  \bibfield  {author} {\bibinfo {author} {\bibfnamefont {L.}~\bibnamefont {Yu}}, \bibinfo {author} {\bibfnamefont {C.}~\bibnamefont {Wang}}, \bibinfo {author} {\bibfnamefont {Y.}~\bibnamefont {Zhang}}, \bibinfo {author} {\bibfnamefont {M.}~\bibnamefont {Sander}}, \bibinfo {author} {\bibfnamefont {S.}~\bibnamefont {Ni}}, \bibinfo {author} {\bibfnamefont {Z.}~\bibnamefont {Lu}}, \bibinfo {author} {\bibfnamefont {S.}~\bibnamefont {Ma}}, \bibinfo {author} {\bibfnamefont {Z.}~\bibnamefont {Wang}}, \bibinfo {author} {\bibfnamefont {Z.}~\bibnamefont {Zhao}}, \bibinfo {author} {\bibfnamefont {H.}~\bibnamefont {Chen}}, \bibinfo {author} {\bibfnamefont {K.}~\bibnamefont {Jiang}}, \bibinfo {author} {\bibfnamefont {Y.}~\bibnamefont {Zhang}}, \bibinfo {author} {\bibfnamefont {H.}~\bibnamefont {Yang}}, \bibinfo {author} {\bibfnamefont {F.}~\bibnamefont {Zhou}}, \bibinfo {author} {\bibfnamefont {X.}~\bibnamefont {Dong}}, \bibinfo {author} {\bibfnamefont {S.~L.}\ \bibnamefont {Johnson}}, \bibinfo {author} {\bibfnamefont
  {M.~J.}\ \bibnamefont {Graf}}, \bibinfo {author} {\bibfnamefont {J.}~\bibnamefont {Hu}}, \bibinfo {author} {\bibfnamefont {H.-J.}\ \bibnamefont {Gao}}, \ and\ \bibinfo {author} {\bibfnamefont {Z.}~\bibnamefont {Zhao}},\ }\bibfield  {title} {\enquote {\bibinfo {title} {Evidence of a hidden flux phase in the topological kagome metal {CsV$_3$Sb$_5$}},}\ }\href@noop {} {\bibfield  {journal} {\bibinfo  {journal} {arXiv:2107.10714}\ } (\bibinfo {year} {2021}{\natexlab{a}})}\BibitemShut {NoStop}%
\bibitem [{\citenamefont {Yu}\ \emph {et~al.}(2021{\natexlab{b}})\citenamefont {Yu}, \citenamefont {Wu}, \citenamefont {Wang}, \citenamefont {Lei}, \citenamefont {Zhuo}, \citenamefont {Ying},\ and\ \citenamefont {Chen}}]{Yu2021b}%
  \BibitemOpen
  \bibfield  {author} {\bibinfo {author} {\bibfnamefont {F.~H.}\ \bibnamefont {Yu}}, \bibinfo {author} {\bibfnamefont {T.}~\bibnamefont {Wu}}, \bibinfo {author} {\bibfnamefont {Z.~Y.}\ \bibnamefont {Wang}}, \bibinfo {author} {\bibfnamefont {B.}~\bibnamefont {Lei}}, \bibinfo {author} {\bibfnamefont {W.~Z.}\ \bibnamefont {Zhuo}}, \bibinfo {author} {\bibfnamefont {J.~J.}\ \bibnamefont {Ying}}, \ and\ \bibinfo {author} {\bibfnamefont {X.~H.}\ \bibnamefont {Chen}},\ }\bibfield  {title} {\enquote {\bibinfo {title} {Concurrence of anomalous {Hall} effect and charge density wave in a superconducting topological kagome metal},}\ }\href@noop {} {\bibfield  {journal} {\bibinfo  {journal} {Phys. Rev. B}\ }\textbf {\bibinfo {volume} {104}},\ \bibinfo {pages} {L041103} (\bibinfo {year} {2021}{\natexlab{b}})}\BibitemShut {NoStop}%
\bibitem [{\citenamefont {Saykin}\ \emph {et~al.}(2023)\citenamefont {Saykin}, \citenamefont {Farhang}, \citenamefont {Kountz}, \citenamefont {Chen}, \citenamefont {Ortiz}, \citenamefont {Shekhar}, \citenamefont {Felser}, \citenamefont {Wilson}, \citenamefont {Thomale}, \citenamefont {Xia} \emph {et~al.}}]{Saykin2023}%
  \BibitemOpen
  \bibfield  {author} {\bibinfo {author} {\bibfnamefont {D.~R.}\ \bibnamefont {Saykin}}, \bibinfo {author} {\bibfnamefont {C.}~\bibnamefont {Farhang}}, \bibinfo {author} {\bibfnamefont {E.~D.}\ \bibnamefont {Kountz}}, \bibinfo {author} {\bibfnamefont {D.}~\bibnamefont {Chen}}, \bibinfo {author} {\bibfnamefont {B.~R.}\ \bibnamefont {Ortiz}}, \bibinfo {author} {\bibfnamefont {C.}~\bibnamefont {Shekhar}}, \bibinfo {author} {\bibfnamefont {C.}~\bibnamefont {Felser}}, \bibinfo {author} {\bibfnamefont {S.~D.}\ \bibnamefont {Wilson}}, \bibinfo {author} {\bibfnamefont {R.}~\bibnamefont {Thomale}}, \bibinfo {author} {\bibfnamefont {J.}~\bibnamefont {Xia}},  \emph {et~al.},\ }\bibfield  {title} {\enquote {\bibinfo {title} {High resolution polar {Kerr} effect studies of {CsV$_3$Sb$_5$}: tests for time-reversal symmetry breaking below the charge-order transition},}\ }\href@noop {} {\bibfield  {journal} {\bibinfo  {journal} {Phys. Rev. Lett.}\ }\textbf {\bibinfo {volume} {131}},\ \bibinfo {pages} {016901} (\bibinfo {year}
  {2023})}\BibitemShut {NoStop}%
\bibitem [{\citenamefont {Farhang}\ \emph {et~al.}(2023)\citenamefont {Farhang}, \citenamefont {Wang}, \citenamefont {Ortiz}, \citenamefont {Wilson},\ and\ \citenamefont {Xia}}]{Farhang2023}%
  \BibitemOpen
  \bibfield  {author} {\bibinfo {author} {\bibfnamefont {C.}~\bibnamefont {Farhang}}, \bibinfo {author} {\bibfnamefont {J.}~\bibnamefont {Wang}}, \bibinfo {author} {\bibfnamefont {B.~R.}\ \bibnamefont {Ortiz}}, \bibinfo {author} {\bibfnamefont {S.~D.}\ \bibnamefont {Wilson}}, \ and\ \bibinfo {author} {\bibfnamefont {J.}~\bibnamefont {Xia}},\ }\bibfield  {title} {\enquote {\bibinfo {title} {Unconventional specular optical rotation in the charge ordered state of kagome metal {CsV$_3$Sb$_5$}},}\ }\href@noop {} {\bibfield  {journal} {\bibinfo  {journal} {Nat. Commun.}\ }\textbf {\bibinfo {volume} {14}},\ \bibinfo {pages} {5326} (\bibinfo {year} {2023})}\BibitemShut {NoStop}%
\bibitem [{\citenamefont {Li}\ \emph {et~al.}(2021)\citenamefont {Li}, \citenamefont {Zhang}, \citenamefont {Yilmaz}, \citenamefont {Pai}, \citenamefont {Marvinney}, \citenamefont {Said}, \citenamefont {Yin}, \citenamefont {Gong}, \citenamefont {Tu}, \citenamefont {Vescovo}, \citenamefont {Nelson}, \citenamefont {Moore}, \citenamefont {Murakami}, \citenamefont {Lei}, \citenamefont {Lee}, \citenamefont {Lawrie},\ and\ \citenamefont {Miao}}]{Li2021}%
  \BibitemOpen
  \bibfield  {author} {\bibinfo {author} {\bibfnamefont {H.}~\bibnamefont {Li}}, \bibinfo {author} {\bibfnamefont {T.~T.}\ \bibnamefont {Zhang}}, \bibinfo {author} {\bibfnamefont {T.}~\bibnamefont {Yilmaz}}, \bibinfo {author} {\bibfnamefont {Y.~Y.}\ \bibnamefont {Pai}}, \bibinfo {author} {\bibfnamefont {C.~E.}\ \bibnamefont {Marvinney}}, \bibinfo {author} {\bibfnamefont {A.}~\bibnamefont {Said}}, \bibinfo {author} {\bibfnamefont {Q.~W.}\ \bibnamefont {Yin}}, \bibinfo {author} {\bibfnamefont {C.~S.}\ \bibnamefont {Gong}}, \bibinfo {author} {\bibfnamefont {Z.~J.}\ \bibnamefont {Tu}}, \bibinfo {author} {\bibfnamefont {E.}~\bibnamefont {Vescovo}}, \bibinfo {author} {\bibfnamefont {C.~S.}\ \bibnamefont {Nelson}}, \bibinfo {author} {\bibfnamefont {R.~G.}\ \bibnamefont {Moore}}, \bibinfo {author} {\bibfnamefont {S.}~\bibnamefont {Murakami}}, \bibinfo {author} {\bibfnamefont {H.~C.}\ \bibnamefont {Lei}}, \bibinfo {author} {\bibfnamefont {H.~N.}\ \bibnamefont {Lee}}, \bibinfo {author} {\bibfnamefont {B.~J.}\
  \bibnamefont {Lawrie}}, \ and\ \bibinfo {author} {\bibfnamefont {H.}~\bibnamefont {Miao}},\ }\bibfield  {title} {\enquote {\bibinfo {title} {Observation of unconventional charge density wave without acoustic phonon anomaly in kagome superconductors {$A$V$_3$Sb$_5$ ($A=\mathrm{Rb}$, Cs)}},}\ }\href {\doibase 10.1103/PhysRevX.11.031050} {\bibfield  {journal} {\bibinfo  {journal} {Phys. Rev. X}\ }\textbf {\bibinfo {volume} {11}},\ \bibinfo {pages} {031050} (\bibinfo {year} {2021})}\BibitemShut {NoStop}%
\bibitem [{\citenamefont {Liang}\ \emph {et~al.}(2021)\citenamefont {Liang}, \citenamefont {Hou}, \citenamefont {Zhang}, \citenamefont {Ma}, \citenamefont {Wu}, \citenamefont {Zhang}, \citenamefont {Yu}, \citenamefont {Ying}, \citenamefont {Jiang}, \citenamefont {Shan}, \citenamefont {Wang},\ and\ \citenamefont {Chen}}]{Liang2021}%
  \BibitemOpen
  \bibfield  {author} {\bibinfo {author} {\bibfnamefont {Z.}~\bibnamefont {Liang}}, \bibinfo {author} {\bibfnamefont {X.}~\bibnamefont {Hou}}, \bibinfo {author} {\bibfnamefont {F.}~\bibnamefont {Zhang}}, \bibinfo {author} {\bibfnamefont {W.}~\bibnamefont {Ma}}, \bibinfo {author} {\bibfnamefont {P.}~\bibnamefont {Wu}}, \bibinfo {author} {\bibfnamefont {Z.}~\bibnamefont {Zhang}}, \bibinfo {author} {\bibfnamefont {F.}~\bibnamefont {Yu}}, \bibinfo {author} {\bibfnamefont {J.-J.}\ \bibnamefont {Ying}}, \bibinfo {author} {\bibfnamefont {K.}~\bibnamefont {Jiang}}, \bibinfo {author} {\bibfnamefont {L.}~\bibnamefont {Shan}}, \bibinfo {author} {\bibfnamefont {Z.}~\bibnamefont {Wang}}, \ and\ \bibinfo {author} {\bibfnamefont {X.-H.}\ \bibnamefont {Chen}},\ }\bibfield  {title} {\enquote {\bibinfo {title} {Three-dimensional charge density wave and surface-dependent vortex-core states in a kagome superconductor {CsV$_3$Sb$_5$}},}\ }\href {\doibase 10.1103/PhysRevX.11.031026} {\bibfield  {journal} {\bibinfo  {journal}
  {Phys. Rev. X}\ }\textbf {\bibinfo {volume} {11}},\ \bibinfo {pages} {031026} (\bibinfo {year} {2021})}\BibitemShut {NoStop}%
\bibitem [{\citenamefont {Kang}\ \emph {et~al.}(2023)\citenamefont {Kang}, \citenamefont {Fang}, \citenamefont {Yoo}, \citenamefont {Ortiz}, \citenamefont {Oey}, \citenamefont {Choi}, \citenamefont {Ryu}, \citenamefont {Kim}, \citenamefont {Jozwiak}, \citenamefont {Bostwick} \emph {et~al.}}]{Kang2023a}%
  \BibitemOpen
  \bibfield  {author} {\bibinfo {author} {\bibfnamefont {M.}~\bibnamefont {Kang}}, \bibinfo {author} {\bibfnamefont {S.}~\bibnamefont {Fang}}, \bibinfo {author} {\bibfnamefont {J.}~\bibnamefont {Yoo}}, \bibinfo {author} {\bibfnamefont {B.~R.}\ \bibnamefont {Ortiz}}, \bibinfo {author} {\bibfnamefont {Y.~M.}\ \bibnamefont {Oey}}, \bibinfo {author} {\bibfnamefont {J.}~\bibnamefont {Choi}}, \bibinfo {author} {\bibfnamefont {S.~H.}\ \bibnamefont {Ryu}}, \bibinfo {author} {\bibfnamefont {J.}~\bibnamefont {Kim}}, \bibinfo {author} {\bibfnamefont {C.}~\bibnamefont {Jozwiak}}, \bibinfo {author} {\bibfnamefont {A.}~\bibnamefont {Bostwick}},  \emph {et~al.},\ }\bibfield  {title} {\enquote {\bibinfo {title} {Charge order landscape and competition with superconductivity in kagome metals},}\ }\href@noop {} {\bibfield  {journal} {\bibinfo  {journal} {Nat. Mater.}\ }\textbf {\bibinfo {volume} {22}},\ \bibinfo {pages} {186} (\bibinfo {year} {2023})}\BibitemShut {NoStop}%
\bibitem [{\citenamefont {Hu}\ \emph {et~al.}(2022{\natexlab{b}})\citenamefont {Hu}, \citenamefont {Wu}, \citenamefont {Ortiz}, \citenamefont {Han}, \citenamefont {Plumb}, \citenamefont {Wilson}, \citenamefont {Schnyder}, \citenamefont {Shi} \emph {et~al.}}]{Hu2022a}%
  \BibitemOpen
  \bibfield  {author} {\bibinfo {author} {\bibfnamefont {Y.}~\bibnamefont {Hu}}, \bibinfo {author} {\bibfnamefont {X.}~\bibnamefont {Wu}}, \bibinfo {author} {\bibfnamefont {B.~R.}\ \bibnamefont {Ortiz}}, \bibinfo {author} {\bibfnamefont {X.}~\bibnamefont {Han}}, \bibinfo {author} {\bibfnamefont {N.~C.}\ \bibnamefont {Plumb}}, \bibinfo {author} {\bibfnamefont {S.~D.}\ \bibnamefont {Wilson}}, \bibinfo {author} {\bibfnamefont {A.~P.}\ \bibnamefont {Schnyder}}, \bibinfo {author} {\bibfnamefont {M.}~\bibnamefont {Shi}},  \emph {et~al.},\ }\bibfield  {title} {\enquote {\bibinfo {title} {Coexistence of trihexagonal and star-of-{David} pattern in the charge density wave of the kagome superconductor {AV$_3$Sb$_5$}},}\ }\href@noop {} {\bibfield  {journal} {\bibinfo  {journal} {Phys. Rev. B}\ }\textbf {\bibinfo {volume} {106}},\ \bibinfo {pages} {L241106} (\bibinfo {year} {2022}{\natexlab{b}})}\BibitemShut {NoStop}%
\bibitem [{\citenamefont {Yu}\ \emph {et~al.}(2021{\natexlab{c}})\citenamefont {Yu}, \citenamefont {Ma}, \citenamefont {Zhuo}, \citenamefont {Liu}, \citenamefont {Wen}, \citenamefont {Lei}, \citenamefont {Ying},\ and\ \citenamefont {Chen}}]{Yu2021}%
  \BibitemOpen
  \bibfield  {author} {\bibinfo {author} {\bibfnamefont {F.}~\bibnamefont {Yu}}, \bibinfo {author} {\bibfnamefont {D.}~\bibnamefont {Ma}}, \bibinfo {author} {\bibfnamefont {W.}~\bibnamefont {Zhuo}}, \bibinfo {author} {\bibfnamefont {S.}~\bibnamefont {Liu}}, \bibinfo {author} {\bibfnamefont {X.}~\bibnamefont {Wen}}, \bibinfo {author} {\bibfnamefont {B.}~\bibnamefont {Lei}}, \bibinfo {author} {\bibfnamefont {J.}~\bibnamefont {Ying}}, \ and\ \bibinfo {author} {\bibfnamefont {X.}~\bibnamefont {Chen}},\ }\bibfield  {title} {\enquote {\bibinfo {title} {Unusual competition of superconductivity and charge-density-wave state in a compressed topological kagome metal},}\ }\href@noop {} {\bibfield  {journal} {\bibinfo  {journal} {Nat. Commun.}\ }\textbf {\bibinfo {volume} {12}},\ \bibinfo {pages} {3645} (\bibinfo {year} {2021}{\natexlab{c}})}\BibitemShut {NoStop}%
\bibitem [{\citenamefont {Fu}\ \emph {et~al.}(2021)\citenamefont {Fu}, \citenamefont {Zhao}, \citenamefont {Chen}, \citenamefont {Yin}, \citenamefont {Tu}, \citenamefont {Gong}, \citenamefont {Xi}, \citenamefont {Zhu}, \citenamefont {Sun}, \citenamefont {Liu},\ and\ \citenamefont {Lei}}]{Fu2021}%
  \BibitemOpen
  \bibfield  {author} {\bibinfo {author} {\bibfnamefont {Y.}~\bibnamefont {Fu}}, \bibinfo {author} {\bibfnamefont {N.}~\bibnamefont {Zhao}}, \bibinfo {author} {\bibfnamefont {Z.}~\bibnamefont {Chen}}, \bibinfo {author} {\bibfnamefont {Q.}~\bibnamefont {Yin}}, \bibinfo {author} {\bibfnamefont {Z.}~\bibnamefont {Tu}}, \bibinfo {author} {\bibfnamefont {C.}~\bibnamefont {Gong}}, \bibinfo {author} {\bibfnamefont {C.}~\bibnamefont {Xi}}, \bibinfo {author} {\bibfnamefont {X.}~\bibnamefont {Zhu}}, \bibinfo {author} {\bibfnamefont {Y.}~\bibnamefont {Sun}}, \bibinfo {author} {\bibfnamefont {K.}~\bibnamefont {Liu}}, \ and\ \bibinfo {author} {\bibfnamefont {H.}~\bibnamefont {Lei}},\ }\bibfield  {title} {\enquote {\bibinfo {title} {Quantum transport evidence of topological band structures of kagome superconductor {CsV$_3$Sb$_5$}},}\ }\href {\doibase 10.1103/PhysRevLett.127.207002} {\bibfield  {journal} {\bibinfo  {journal} {Phys. Rev. Lett.}\ }\textbf {\bibinfo {volume} {127}},\ \bibinfo {pages} {207002} (\bibinfo {year}
  {2021})}\BibitemShut {NoStop}%
\bibitem [{\citenamefont {Gan}\ \emph {et~al.}(2021)\citenamefont {Gan}, \citenamefont {Xia}, \citenamefont {Zhang}, \citenamefont {Yang}, \citenamefont {Mi}, \citenamefont {Wang}, \citenamefont {Chai}, \citenamefont {Guo}, \citenamefont {Zhou},\ and\ \citenamefont {He}}]{Gan2021}%
  \BibitemOpen
  \bibfield  {author} {\bibinfo {author} {\bibfnamefont {Y.}~\bibnamefont {Gan}}, \bibinfo {author} {\bibfnamefont {W.}~\bibnamefont {Xia}}, \bibinfo {author} {\bibfnamefont {L.}~\bibnamefont {Zhang}}, \bibinfo {author} {\bibfnamefont {K.}~\bibnamefont {Yang}}, \bibinfo {author} {\bibfnamefont {X.}~\bibnamefont {Mi}}, \bibinfo {author} {\bibfnamefont {A.}~\bibnamefont {Wang}}, \bibinfo {author} {\bibfnamefont {Y.}~\bibnamefont {Chai}}, \bibinfo {author} {\bibfnamefont {Y.}~\bibnamefont {Guo}}, \bibinfo {author} {\bibfnamefont {X.}~\bibnamefont {Zhou}}, \ and\ \bibinfo {author} {\bibfnamefont {M.}~\bibnamefont {He}},\ }\bibfield  {title} {\enquote {\bibinfo {title} {Magneto-{Seebeck} effect and ambipolar {Nernst} effect in the {CsV$_3$Sb$_5$} superconductor},}\ }\href@noop {} {\bibfield  {journal} {\bibinfo  {journal} {Phys. Rev. B}\ }\textbf {\bibinfo {volume} {104}},\ \bibinfo {pages} {L180508} (\bibinfo {year} {2021})}\BibitemShut {NoStop}%
\bibitem [{\citenamefont {Zhang}\ \emph {et~al.}(2022)\citenamefont {Zhang}, \citenamefont {Wang}, \citenamefont {Tsang}, \citenamefont {Liu}, \citenamefont {Xie}, \citenamefont {Yu}, \citenamefont {Lai},\ and\ \citenamefont {Goh}}]{Zhang2022}%
  \BibitemOpen
  \bibfield  {author} {\bibinfo {author} {\bibfnamefont {W.}~\bibnamefont {Zhang}}, \bibinfo {author} {\bibfnamefont {L.}~\bibnamefont {Wang}}, \bibinfo {author} {\bibfnamefont {C.~W.}\ \bibnamefont {Tsang}}, \bibinfo {author} {\bibfnamefont {X.}~\bibnamefont {Liu}}, \bibinfo {author} {\bibfnamefont {J.}~\bibnamefont {Xie}}, \bibinfo {author} {\bibfnamefont {W.~C.}\ \bibnamefont {Yu}}, \bibinfo {author} {\bibfnamefont {K.~T.}\ \bibnamefont {Lai}}, \ and\ \bibinfo {author} {\bibfnamefont {S.~K.}\ \bibnamefont {Goh}},\ }\bibfield  {title} {\enquote {\bibinfo {title} {Emergence of large quantum oscillation frequencies in thin flakes of the kagome superconductor {CsV$_3$Sb$_5$}},}\ }\href@noop {} {\bibfield  {journal} {\bibinfo  {journal} {Phys. Rev. B}\ }\textbf {\bibinfo {volume} {106}},\ \bibinfo {pages} {195103} (\bibinfo {year} {2022})}\BibitemShut {NoStop}%
\bibitem [{\citenamefont {Chen}\ \emph {et~al.}(2022)\citenamefont {Chen}, \citenamefont {He}, \citenamefont {Yao}, \citenamefont {Pan}, \citenamefont {Lin}, \citenamefont {Schnelle}, \citenamefont {Sun}, \citenamefont {Gooth}, \citenamefont {Taillefer},\ and\ \citenamefont {Felser}}]{Chen2022}%
  \BibitemOpen
  \bibfield  {author} {\bibinfo {author} {\bibfnamefont {D.}~\bibnamefont {Chen}}, \bibinfo {author} {\bibfnamefont {B.}~\bibnamefont {He}}, \bibinfo {author} {\bibfnamefont {M.}~\bibnamefont {Yao}}, \bibinfo {author} {\bibfnamefont {Y.}~\bibnamefont {Pan}}, \bibinfo {author} {\bibfnamefont {H.}~\bibnamefont {Lin}}, \bibinfo {author} {\bibfnamefont {W.}~\bibnamefont {Schnelle}}, \bibinfo {author} {\bibfnamefont {Y.}~\bibnamefont {Sun}}, \bibinfo {author} {\bibfnamefont {J.}~\bibnamefont {Gooth}}, \bibinfo {author} {\bibfnamefont {L.}~\bibnamefont {Taillefer}}, \ and\ \bibinfo {author} {\bibfnamefont {C.}~\bibnamefont {Felser}},\ }\bibfield  {title} {\enquote {\bibinfo {title} {Anomalous thermoelectric effects and quantum oscillations in the kagome metal {CsV$_3$Sb$_5$}},}\ }\href@noop {} {\bibfield  {journal} {\bibinfo  {journal} {Phys. Rev. B}\ }\textbf {\bibinfo {volume} {105}},\ \bibinfo {pages} {L201109} (\bibinfo {year} {2022})}\BibitemShut {NoStop}%
\bibitem [{\citenamefont {Huang}\ \emph {et~al.}(2022)\citenamefont {Huang}, \citenamefont {Guo}, \citenamefont {Putzke}, \citenamefont {Gutierrez-Amigo}, \citenamefont {Sun}, \citenamefont {Vergniory}, \citenamefont {Errea}, \citenamefont {Chen}, \citenamefont {Felser},\ and\ \citenamefont {Moll}}]{Huang2022}%
  \BibitemOpen
  \bibfield  {author} {\bibinfo {author} {\bibfnamefont {X.}~\bibnamefont {Huang}}, \bibinfo {author} {\bibfnamefont {C.}~\bibnamefont {Guo}}, \bibinfo {author} {\bibfnamefont {C.}~\bibnamefont {Putzke}}, \bibinfo {author} {\bibfnamefont {M.}~\bibnamefont {Gutierrez-Amigo}}, \bibinfo {author} {\bibfnamefont {Y.}~\bibnamefont {Sun}}, \bibinfo {author} {\bibfnamefont {M.~G.}\ \bibnamefont {Vergniory}}, \bibinfo {author} {\bibfnamefont {I.}~\bibnamefont {Errea}}, \bibinfo {author} {\bibfnamefont {D.}~\bibnamefont {Chen}}, \bibinfo {author} {\bibfnamefont {C.}~\bibnamefont {Felser}}, \ and\ \bibinfo {author} {\bibfnamefont {P.~J.~W.}\ \bibnamefont {Moll}},\ }\bibfield  {title} {\enquote {\bibinfo {title} {Three-dimensional {Fermi} surfaces from charge order in layered {CsV$_3$Sb$_5$}},}\ }\href {\doibase 10.1103/PhysRevB.106.064510} {\bibfield  {journal} {\bibinfo  {journal} {Phys. Rev. B}\ }\textbf {\bibinfo {volume} {106}},\ \bibinfo {pages} {064510} (\bibinfo {year} {2022})}\BibitemShut {NoStop}%
\bibitem [{\citenamefont {Shrestha}\ \emph {et~al.}(2022)\citenamefont {Shrestha}, \citenamefont {Chapai}, \citenamefont {Pokharel}, \citenamefont {Miertschin}, \citenamefont {Nguyen}, \citenamefont {Zhou}, \citenamefont {Chung}, \citenamefont {Kanatzidis}, \citenamefont {Mitchell}, \citenamefont {Welp}, \citenamefont {Popovi\ifmmode~\acute{c}\else \'{c}\fi{}}, \citenamefont {Graf}, \citenamefont {Lorenz},\ and\ \citenamefont {Kwok}}]{Shrestha2022}%
  \BibitemOpen
  \bibfield  {author} {\bibinfo {author} {\bibfnamefont {K.}~\bibnamefont {Shrestha}}, \bibinfo {author} {\bibfnamefont {R.}~\bibnamefont {Chapai}}, \bibinfo {author} {\bibfnamefont {B.~K.}\ \bibnamefont {Pokharel}}, \bibinfo {author} {\bibfnamefont {D.}~\bibnamefont {Miertschin}}, \bibinfo {author} {\bibfnamefont {T.}~\bibnamefont {Nguyen}}, \bibinfo {author} {\bibfnamefont {X.}~\bibnamefont {Zhou}}, \bibinfo {author} {\bibfnamefont {D.~Y.}\ \bibnamefont {Chung}}, \bibinfo {author} {\bibfnamefont {M.~G.}\ \bibnamefont {Kanatzidis}}, \bibinfo {author} {\bibfnamefont {J.~F.}\ \bibnamefont {Mitchell}}, \bibinfo {author} {\bibfnamefont {U.}~\bibnamefont {Welp}}, \bibinfo {author} {\bibfnamefont {D.}~\bibnamefont {Popovi\ifmmode~\acute{c}\else \'{c}\fi{}}}, \bibinfo {author} {\bibfnamefont {D.~E.}\ \bibnamefont {Graf}}, \bibinfo {author} {\bibfnamefont {B.}~\bibnamefont {Lorenz}}, \ and\ \bibinfo {author} {\bibfnamefont {W.~K.}\ \bibnamefont {Kwok}},\ }\bibfield  {title} {\enquote {\bibinfo {title} {Nontrivial
  {Fermi} surface topology of the kagome superconductor {CsV$_3$Sb$_5$} probed by {de Haas--van Alphen} oscillations},}\ }\href {\doibase 10.1103/PhysRevB.105.024508} {\bibfield  {journal} {\bibinfo  {journal} {Phys. Rev. B}\ }\textbf {\bibinfo {volume} {105}},\ \bibinfo {pages} {024508} (\bibinfo {year} {2022})}\BibitemShut {NoStop}%
\bibitem [{\citenamefont {Broyles}\ \emph {et~al.}(2022)\citenamefont {Broyles}, \citenamefont {Graf}, \citenamefont {Yang}, \citenamefont {Dong}, \citenamefont {Gao},\ and\ \citenamefont {Ran}}]{Broyles2022}%
  \BibitemOpen
  \bibfield  {author} {\bibinfo {author} {\bibfnamefont {C.}~\bibnamefont {Broyles}}, \bibinfo {author} {\bibfnamefont {D.}~\bibnamefont {Graf}}, \bibinfo {author} {\bibfnamefont {H.}~\bibnamefont {Yang}}, \bibinfo {author} {\bibfnamefont {X.}~\bibnamefont {Dong}}, \bibinfo {author} {\bibfnamefont {H.}~\bibnamefont {Gao}}, \ and\ \bibinfo {author} {\bibfnamefont {S.}~\bibnamefont {Ran}},\ }\bibfield  {title} {\enquote {\bibinfo {title} {Effect of the interlayer ordering on the {Fermi} surface of kagome superconductor {CsV$_3$Sb$_5$} revealed by quantum oscillations},}\ }\href@noop {} {\bibfield  {journal} {\bibinfo  {journal} {Phys. Rev. Lett.}\ }\textbf {\bibinfo {volume} {129}},\ \bibinfo {pages} {157001} (\bibinfo {year} {2022})}\BibitemShut {NoStop}%
\bibitem [{\citenamefont {Chapai}\ \emph {et~al.}(2023)\citenamefont {Chapai}, \citenamefont {Leroux}, \citenamefont {Oliviero}, \citenamefont {Vignolles}, \citenamefont {Bruyant}, \citenamefont {Smylie}, \citenamefont {Chung}, \citenamefont {Kanatzidis}, \citenamefont {Kwok}, \citenamefont {Mitchell} \emph {et~al.}}]{Chapai2023}%
  \BibitemOpen
  \bibfield  {author} {\bibinfo {author} {\bibfnamefont {R.}~\bibnamefont {Chapai}}, \bibinfo {author} {\bibfnamefont {M.}~\bibnamefont {Leroux}}, \bibinfo {author} {\bibfnamefont {V.}~\bibnamefont {Oliviero}}, \bibinfo {author} {\bibfnamefont {D.}~\bibnamefont {Vignolles}}, \bibinfo {author} {\bibfnamefont {N.}~\bibnamefont {Bruyant}}, \bibinfo {author} {\bibfnamefont {M.}~\bibnamefont {Smylie}}, \bibinfo {author} {\bibfnamefont {D.}~\bibnamefont {Chung}}, \bibinfo {author} {\bibfnamefont {M.}~\bibnamefont {Kanatzidis}}, \bibinfo {author} {\bibfnamefont {W.-K.}\ \bibnamefont {Kwok}}, \bibinfo {author} {\bibfnamefont {J.}~\bibnamefont {Mitchell}},  \emph {et~al.},\ }\bibfield  {title} {\enquote {\bibinfo {title} {Magnetic breakdown and topology in the kagome superconductor {CsV$_3$Sb$_5$} under high magnetic field},}\ }\href@noop {} {\bibfield  {journal} {\bibinfo  {journal} {Phys. Rev. Lett.}\ }\textbf {\bibinfo {volume} {130}},\ \bibinfo {pages} {126401} (\bibinfo {year} {2023})}\BibitemShut {NoStop}%
\bibitem [{\citenamefont {Chen}\ \emph {et~al.}(2021)\citenamefont {Chen}, \citenamefont {Wang}, \citenamefont {Yin}, \citenamefont {Gu}, \citenamefont {Jiang}, \citenamefont {Tu}, \citenamefont {Gong}, \citenamefont {Uwatoko}, \citenamefont {Sun}, \citenamefont {Lei}, \citenamefont {Hu},\ and\ \citenamefont {Cheng}}]{Chen2021a}%
  \BibitemOpen
  \bibfield  {author} {\bibinfo {author} {\bibfnamefont {K.~Y.}\ \bibnamefont {Chen}}, \bibinfo {author} {\bibfnamefont {N.~N.}\ \bibnamefont {Wang}}, \bibinfo {author} {\bibfnamefont {Q.~W.}\ \bibnamefont {Yin}}, \bibinfo {author} {\bibfnamefont {Y.~H.}\ \bibnamefont {Gu}}, \bibinfo {author} {\bibfnamefont {K.}~\bibnamefont {Jiang}}, \bibinfo {author} {\bibfnamefont {Z.~J.}\ \bibnamefont {Tu}}, \bibinfo {author} {\bibfnamefont {C.~S.}\ \bibnamefont {Gong}}, \bibinfo {author} {\bibfnamefont {Y.}~\bibnamefont {Uwatoko}}, \bibinfo {author} {\bibfnamefont {J.~P.}\ \bibnamefont {Sun}}, \bibinfo {author} {\bibfnamefont {H.~C.}\ \bibnamefont {Lei}}, \bibinfo {author} {\bibfnamefont {J.~P.}\ \bibnamefont {Hu}}, \ and\ \bibinfo {author} {\bibfnamefont {J.-G.}\ \bibnamefont {Cheng}},\ }\bibfield  {title} {\enquote {\bibinfo {title} {Double superconducting dome and triple enhancement of {${T}_{c}$} in the kagome superconductor {CsV$_3$Sb$_5$} under high pressure},}\ }\href {\doibase 10.1103/PhysRevLett.126.247001}
  {\bibfield  {journal} {\bibinfo  {journal} {Phys. Rev. Lett.}\ }\textbf {\bibinfo {volume} {126}},\ \bibinfo {pages} {247001} (\bibinfo {year} {2021})}\BibitemShut {NoStop}%
\bibitem [{\citenamefont {Tazai}\ \emph {et~al.}(2022)\citenamefont {Tazai}, \citenamefont {Yamakawa}, \citenamefont {Onari},\ and\ \citenamefont {Kontani}}]{Tazai2022}%
  \BibitemOpen
  \bibfield  {author} {\bibinfo {author} {\bibfnamefont {R.}~\bibnamefont {Tazai}}, \bibinfo {author} {\bibfnamefont {Y.}~\bibnamefont {Yamakawa}}, \bibinfo {author} {\bibfnamefont {S.}~\bibnamefont {Onari}}, \ and\ \bibinfo {author} {\bibfnamefont {H.}~\bibnamefont {Kontani}},\ }\bibfield  {title} {\enquote {\bibinfo {title} {Mechanism of exotic density-wave and beyond-migdal unconventional superconductivity in kagome metal {AV$_3$Sb$_5$} ({A= K, Rb, Cs})},}\ }\href@noop {} {\bibfield  {journal} {\bibinfo  {journal} {Sci. Adv.}\ }\textbf {\bibinfo {volume} {8}},\ \bibinfo {pages} {eabl4108} (\bibinfo {year} {2022})}\BibitemShut {NoStop}%
\bibitem [{\citenamefont {Shishido}\ \emph {et~al.}(2005)\citenamefont {Shishido}, \citenamefont {Settai}, \citenamefont {Harima},\ and\ \citenamefont {{\=O}nuki}}]{Shishido2005}%
  \BibitemOpen
  \bibfield  {author} {\bibinfo {author} {\bibfnamefont {H.}~\bibnamefont {Shishido}}, \bibinfo {author} {\bibfnamefont {R.}~\bibnamefont {Settai}}, \bibinfo {author} {\bibfnamefont {H.}~\bibnamefont {Harima}}, \ and\ \bibinfo {author} {\bibfnamefont {Y.}~\bibnamefont {{\=O}nuki}},\ }\bibfield  {title} {\enquote {\bibinfo {title} {A drastic change of the {Fermi} surface at a critical pressure in {CeRhIn$_5$}: {dHvA} study under pressure},}\ }\href@noop {} {\bibfield  {journal} {\bibinfo  {journal} {J. Phys. Soc. Jpn.}\ }\textbf {\bibinfo {volume} {74}},\ \bibinfo {pages} {1103} (\bibinfo {year} {2005})}\BibitemShut {NoStop}%
\bibitem [{\citenamefont {Shishido}\ \emph {et~al.}(2010)\citenamefont {Shishido}, \citenamefont {Bangura}, \citenamefont {Coldea}, \citenamefont {Tonegawa}, \citenamefont {Hashimoto}, \citenamefont {Kasahara}, \citenamefont {Ikeda}, \citenamefont {Terashima}, \citenamefont {Settai}, \citenamefont {{\=O}nuki} \emph {et~al.}}]{Shishido2010}%
  \BibitemOpen
  \bibfield  {author} {\bibinfo {author} {\bibfnamefont {H.}~\bibnamefont {Shishido}}, \bibinfo {author} {\bibfnamefont {A.}~\bibnamefont {Bangura}}, \bibinfo {author} {\bibfnamefont {A.}~\bibnamefont {Coldea}}, \bibinfo {author} {\bibfnamefont {S.}~\bibnamefont {Tonegawa}}, \bibinfo {author} {\bibfnamefont {K.}~\bibnamefont {Hashimoto}}, \bibinfo {author} {\bibfnamefont {S.}~\bibnamefont {Kasahara}}, \bibinfo {author} {\bibfnamefont {H.}~\bibnamefont {Ikeda}}, \bibinfo {author} {\bibfnamefont {T.}~\bibnamefont {Terashima}}, \bibinfo {author} {\bibfnamefont {R.}~\bibnamefont {Settai}}, \bibinfo {author} {\bibfnamefont {Y.}~\bibnamefont {{\=O}nuki}},  \emph {et~al.},\ }\bibfield  {title} {\enquote {\bibinfo {title} {Evolution of the {Fermi} surface of {BaFe$_2$(As$_{1-x}$P$_x$)$_2$} on entering the superconducting dome},}\ }\href@noop {} {\bibfield  {journal} {\bibinfo  {journal} {Phys. Rev. Lett.}\ }\textbf {\bibinfo {volume} {104}},\ \bibinfo {pages} {057008} (\bibinfo {year} {2010})}\BibitemShut {NoStop}%
\bibitem [{\citenamefont {Ramshaw}\ \emph {et~al.}(2015)\citenamefont {Ramshaw}, \citenamefont {Sebastian}, \citenamefont {McDonald}, \citenamefont {Day}, \citenamefont {Tan}, \citenamefont {Zhu}, \citenamefont {Betts}, \citenamefont {Liang}, \citenamefont {Bonn}, \citenamefont {Hardy} \emph {et~al.}}]{Ramshaw2015}%
  \BibitemOpen
  \bibfield  {author} {\bibinfo {author} {\bibfnamefont {B.}~\bibnamefont {Ramshaw}}, \bibinfo {author} {\bibfnamefont {S.}~\bibnamefont {Sebastian}}, \bibinfo {author} {\bibfnamefont {R.}~\bibnamefont {McDonald}}, \bibinfo {author} {\bibfnamefont {J.}~\bibnamefont {Day}}, \bibinfo {author} {\bibfnamefont {B.}~\bibnamefont {Tan}}, \bibinfo {author} {\bibfnamefont {Z.}~\bibnamefont {Zhu}}, \bibinfo {author} {\bibfnamefont {J.}~\bibnamefont {Betts}}, \bibinfo {author} {\bibfnamefont {R.}~\bibnamefont {Liang}}, \bibinfo {author} {\bibfnamefont {D.}~\bibnamefont {Bonn}}, \bibinfo {author} {\bibfnamefont {W.}~\bibnamefont {Hardy}},  \emph {et~al.},\ }\bibfield  {title} {\enquote {\bibinfo {title} {Quasiparticle mass enhancement approaching optimal doping in a high-{$T_c$} superconductor},}\ }\href@noop {} {\bibfield  {journal} {\bibinfo  {journal} {Science}\ }\textbf {\bibinfo {volume} {348}},\ \bibinfo {pages} {317} (\bibinfo {year} {2015})}\BibitemShut {NoStop}%
\bibitem [{\citenamefont {Zhang}\ \emph {et~al.}(2023)\citenamefont {Zhang}, \citenamefont {Liu}, \citenamefont {Wang}, \citenamefont {Tsang}, \citenamefont {Wang}, \citenamefont {Lam}, \citenamefont {Wang}, \citenamefont {Xie}, \citenamefont {Zhou}, \citenamefont {Zhao} \emph {et~al.}}]{Zhang2023}%
  \BibitemOpen
  \bibfield  {author} {\bibinfo {author} {\bibfnamefont {W.}~\bibnamefont {Zhang}}, \bibinfo {author} {\bibfnamefont {X.}~\bibnamefont {Liu}}, \bibinfo {author} {\bibfnamefont {L.}~\bibnamefont {Wang}}, \bibinfo {author} {\bibfnamefont {C.~W.}\ \bibnamefont {Tsang}}, \bibinfo {author} {\bibfnamefont {Z.}~\bibnamefont {Wang}}, \bibinfo {author} {\bibfnamefont {S.~T.}\ \bibnamefont {Lam}}, \bibinfo {author} {\bibfnamefont {W.}~\bibnamefont {Wang}}, \bibinfo {author} {\bibfnamefont {J.}~\bibnamefont {Xie}}, \bibinfo {author} {\bibfnamefont {X.}~\bibnamefont {Zhou}}, \bibinfo {author} {\bibfnamefont {Y.}~\bibnamefont {Zhao}},  \emph {et~al.},\ }\bibfield  {title} {\enquote {\bibinfo {title} {Nodeless superconductivity in kagome metal {CsV$_3$Sb$_5$} with and without time reversal symmetry breaking},}\ }\href@noop {} {\bibfield  {journal} {\bibinfo  {journal} {Nano Lett.}\ }\textbf {\bibinfo {volume} {23}},\ \bibinfo {pages} {872} (\bibinfo {year} {2023})}\BibitemShut {NoStop}%
\bibitem [{\citenamefont {Tsirlin}\ \emph {et~al.}(2022)\citenamefont {Tsirlin}, \citenamefont {Fertey}, \citenamefont {Ortiz}, \citenamefont {Klis}, \citenamefont {Merkl}, \citenamefont {Dressel}, \citenamefont {Wilson},\ and\ \citenamefont {Uykur}}]{Tsirlin2022}%
  \BibitemOpen
  \bibfield  {author} {\bibinfo {author} {\bibfnamefont {A.}~\bibnamefont {Tsirlin}}, \bibinfo {author} {\bibfnamefont {P.}~\bibnamefont {Fertey}}, \bibinfo {author} {\bibfnamefont {B.~R.}\ \bibnamefont {Ortiz}}, \bibinfo {author} {\bibfnamefont {B.}~\bibnamefont {Klis}}, \bibinfo {author} {\bibfnamefont {V.}~\bibnamefont {Merkl}}, \bibinfo {author} {\bibfnamefont {M.}~\bibnamefont {Dressel}}, \bibinfo {author} {\bibfnamefont {S.}~\bibnamefont {Wilson}}, \ and\ \bibinfo {author} {\bibfnamefont {E.}~\bibnamefont {Uykur}},\ }\bibfield  {title} {\enquote {\bibinfo {title} {Role of {Sb} in the superconducting kagome metal {CsV$_3$Sb$_5$} revealed by its anisotropic compression},}\ }\href@noop {} {\bibfield  {journal} {\bibinfo  {journal} {SciPost Phys.}\ }\textbf {\bibinfo {volume} {12}},\ \bibinfo {pages} {049} (\bibinfo {year} {2022})}\BibitemShut {NoStop}%
\bibitem [{\citenamefont {Wang}\ \emph {et~al.}(2022)\citenamefont {Wang}, \citenamefont {Liu}, \citenamefont {Jeon}, \citenamefont {Jia},\ and\ \citenamefont {Cho}}]{Wang2022}%
  \BibitemOpen
  \bibfield  {author} {\bibinfo {author} {\bibfnamefont {C.}~\bibnamefont {Wang}}, \bibinfo {author} {\bibfnamefont {S.}~\bibnamefont {Liu}}, \bibinfo {author} {\bibfnamefont {H.}~\bibnamefont {Jeon}}, \bibinfo {author} {\bibfnamefont {Y.}~\bibnamefont {Jia}}, \ and\ \bibinfo {author} {\bibfnamefont {J.-H.}\ \bibnamefont {Cho}},\ }\bibfield  {title} {\enquote {\bibinfo {title} {Charge density wave and superconductivity in the kagome metal {CsV$_3$Sb$_5$} around a pressure-induced quantum critical point},}\ }\href@noop {} {\bibfield  {journal} {\bibinfo  {journal} {Phys. Rev. Mater.}\ }\textbf {\bibinfo {volume} {6}},\ \bibinfo {pages} {094801} (\bibinfo {year} {2022})}\BibitemShut {NoStop}%
\bibitem [{\citenamefont {Wenzel}\ \emph {et~al.}(2023)\citenamefont {Wenzel}, \citenamefont {Tsirlin}, \citenamefont {Capitani}, \citenamefont {Chan}, \citenamefont {Ortiz}, \citenamefont {Wilson}, \citenamefont {Dressel},\ and\ \citenamefont {Uykur}}]{Wenzel2023}%
  \BibitemOpen
  \bibfield  {author} {\bibinfo {author} {\bibfnamefont {M.}~\bibnamefont {Wenzel}}, \bibinfo {author} {\bibfnamefont {A.~A.}\ \bibnamefont {Tsirlin}}, \bibinfo {author} {\bibfnamefont {F.}~\bibnamefont {Capitani}}, \bibinfo {author} {\bibfnamefont {Y.~T.}\ \bibnamefont {Chan}}, \bibinfo {author} {\bibfnamefont {B.~R.}\ \bibnamefont {Ortiz}}, \bibinfo {author} {\bibfnamefont {S.~D.}\ \bibnamefont {Wilson}}, \bibinfo {author} {\bibfnamefont {M.}~\bibnamefont {Dressel}}, \ and\ \bibinfo {author} {\bibfnamefont {E.}~\bibnamefont {Uykur}},\ }\bibfield  {title} {\enquote {\bibinfo {title} {Pressure evolution of electron dynamics in the superconducting kagome metal {CsV$_3$Sb$_5$}},}\ }\href@noop {} {\bibfield  {journal} {\bibinfo  {journal} {npj Quant. Mater.}\ }\textbf {\bibinfo {volume} {8}},\ \bibinfo {pages} {45} (\bibinfo {year} {2023})}\BibitemShut {NoStop}%
\bibitem [{\citenamefont {Zhong}\ \emph {et~al.}(2023)\citenamefont {Zhong}, \citenamefont {Li}, \citenamefont {Liu}, \citenamefont {Dong}, \citenamefont {Aido}, \citenamefont {Arai}, \citenamefont {Li}, \citenamefont {Zhang}, \citenamefont {Shi}, \citenamefont {Wang} \emph {et~al.}}]{Zhong2023}%
  \BibitemOpen
  \bibfield  {author} {\bibinfo {author} {\bibfnamefont {Y.}~\bibnamefont {Zhong}}, \bibinfo {author} {\bibfnamefont {S.}~\bibnamefont {Li}}, \bibinfo {author} {\bibfnamefont {H.}~\bibnamefont {Liu}}, \bibinfo {author} {\bibfnamefont {Y.}~\bibnamefont {Dong}}, \bibinfo {author} {\bibfnamefont {K.}~\bibnamefont {Aido}}, \bibinfo {author} {\bibfnamefont {Y.}~\bibnamefont {Arai}}, \bibinfo {author} {\bibfnamefont {H.}~\bibnamefont {Li}}, \bibinfo {author} {\bibfnamefont {W.}~\bibnamefont {Zhang}}, \bibinfo {author} {\bibfnamefont {Y.}~\bibnamefont {Shi}}, \bibinfo {author} {\bibfnamefont {Z.}~\bibnamefont {Wang}},  \emph {et~al.},\ }\bibfield  {title} {\enquote {\bibinfo {title} {Testing electron--phonon coupling for the superconductivity in kagome metal {CsV$_3$Sb$_5$}},}\ }\href@noop {} {\bibfield  {journal} {\bibinfo  {journal} {Nat. Commun.}\ }\textbf {\bibinfo {volume} {14}},\ \bibinfo {pages} {1945} (\bibinfo {year} {2023})}\BibitemShut {NoStop}%
\bibitem [{\citenamefont {Wang}\ \emph {et~al.}(2023{\natexlab{c}})\citenamefont {Wang}, \citenamefont {Jia}, \citenamefont {Zhang},\ and\ \citenamefont {Cho}}]{Wang2023c}%
  \BibitemOpen
  \bibfield  {author} {\bibinfo {author} {\bibfnamefont {C.}~\bibnamefont {Wang}}, \bibinfo {author} {\bibfnamefont {Y.}~\bibnamefont {Jia}}, \bibinfo {author} {\bibfnamefont {Z.}~\bibnamefont {Zhang}}, \ and\ \bibinfo {author} {\bibfnamefont {J.-H.}\ \bibnamefont {Cho}},\ }\bibfield  {title} {\enquote {\bibinfo {title} {Phonon-mediated $s$-wave superconductivity in the kagome metal {CsV$_3$Sb$_5$} under pressure},}\ }\href {\doibase 10.1103/PhysRevB.108.L060503} {\bibfield  {journal} {\bibinfo  {journal} {Phys. Rev. B}\ }\textbf {\bibinfo {volume} {108}},\ \bibinfo {pages} {L060503} (\bibinfo {year} {2023}{\natexlab{c}})}\BibitemShut {NoStop}%
\bibitem [{\citenamefont {Sipos}\ \emph {et~al.}(2008)\citenamefont {Sipos}, \citenamefont {Kusmartseva}, \citenamefont {Akrap}, \citenamefont {Berger}, \citenamefont {Forr{\'o}},\ and\ \citenamefont {Tuti{\v{s}}}}]{Sipos2008}%
  \BibitemOpen
  \bibfield  {author} {\bibinfo {author} {\bibfnamefont {B.}~\bibnamefont {Sipos}}, \bibinfo {author} {\bibfnamefont {A.~F.}\ \bibnamefont {Kusmartseva}}, \bibinfo {author} {\bibfnamefont {A.}~\bibnamefont {Akrap}}, \bibinfo {author} {\bibfnamefont {H.}~\bibnamefont {Berger}}, \bibinfo {author} {\bibfnamefont {L.}~\bibnamefont {Forr{\'o}}}, \ and\ \bibinfo {author} {\bibfnamefont {E.}~\bibnamefont {Tuti{\v{s}}}},\ }\bibfield  {title} {\enquote {\bibinfo {title} {From {Mott} state to superconductivity in {1T-TaS$_2$}},}\ }\href@noop {} {\bibfield  {journal} {\bibinfo  {journal} {Nat. Mater.}\ }\textbf {\bibinfo {volume} {7}},\ \bibinfo {pages} {960} (\bibinfo {year} {2008})}\BibitemShut {NoStop}%
\bibitem [{\citenamefont {Xie}\ \emph {et~al.}(2021)\citenamefont {Xie}, \citenamefont {Liu}, \citenamefont {Zhang}, \citenamefont {Wong}, \citenamefont {Zhou}, \citenamefont {Zhao}, \citenamefont {Wang}, \citenamefont {Lai},\ and\ \citenamefont {Goh}}]{Xie2021}%
  \BibitemOpen
  \bibfield  {author} {\bibinfo {author} {\bibfnamefont {J.}~\bibnamefont {Xie}}, \bibinfo {author} {\bibfnamefont {X.}~\bibnamefont {Liu}}, \bibinfo {author} {\bibfnamefont {W.}~\bibnamefont {Zhang}}, \bibinfo {author} {\bibfnamefont {S.~M.}\ \bibnamefont {Wong}}, \bibinfo {author} {\bibfnamefont {X.}~\bibnamefont {Zhou}}, \bibinfo {author} {\bibfnamefont {Y.}~\bibnamefont {Zhao}}, \bibinfo {author} {\bibfnamefont {S.}~\bibnamefont {Wang}}, \bibinfo {author} {\bibfnamefont {K.~T.}\ \bibnamefont {Lai}}, \ and\ \bibinfo {author} {\bibfnamefont {S.~K.}\ \bibnamefont {Goh}},\ }\bibfield  {title} {\enquote {\bibinfo {title} {Fragile pressure-induced magnetism in {FeSe} superconductors with a thickness reduction},}\ }\href@noop {} {\bibfield  {journal} {\bibinfo  {journal} {Nano Lett.}\ }\textbf {\bibinfo {volume} {21}},\ \bibinfo {pages} {9310} (\bibinfo {year} {2021})}\BibitemShut {NoStop}%
\bibitem [{\citenamefont {Ku}\ \emph {et~al.}(2022)\citenamefont {Ku}, \citenamefont {Liu}, \citenamefont {Xie}, \citenamefont {Zhang}, \citenamefont {Lam}, \citenamefont {Chen}, \citenamefont {Zhou}, \citenamefont {Zhao}, \citenamefont {Wang}, \citenamefont {Yang} \emph {et~al.}}]{Ku2022}%
  \BibitemOpen
  \bibfield  {author} {\bibinfo {author} {\bibfnamefont {C.-h.}\ \bibnamefont {Ku}}, \bibinfo {author} {\bibfnamefont {X.}~\bibnamefont {Liu}}, \bibinfo {author} {\bibfnamefont {J.}~\bibnamefont {Xie}}, \bibinfo {author} {\bibfnamefont {W.}~\bibnamefont {Zhang}}, \bibinfo {author} {\bibfnamefont {S.~T.}\ \bibnamefont {Lam}}, \bibinfo {author} {\bibfnamefont {Y.}~\bibnamefont {Chen}}, \bibinfo {author} {\bibfnamefont {X.}~\bibnamefont {Zhou}}, \bibinfo {author} {\bibfnamefont {Y.}~\bibnamefont {Zhao}}, \bibinfo {author} {\bibfnamefont {S.}~\bibnamefont {Wang}}, \bibinfo {author} {\bibfnamefont {S.}~\bibnamefont {Yang}},  \emph {et~al.},\ }\bibfield  {title} {\enquote {\bibinfo {title} {Patterned diamond anvils prepared via laser writing for electrical transport measurements of thin quantum materials under pressure},}\ }\href@noop {} {\bibfield  {journal} {\bibinfo  {journal} {Rev. Sci. Instrum.}\ }\textbf {\bibinfo {volume} {93}},\ \bibinfo {pages} {083912} (\bibinfo {year} {2022})}\BibitemShut {NoStop}%
\bibitem [{\citenamefont {Schwarz}\ and\ \citenamefont {Blaha}(2003)}]{schwarz2003solid}%
  \BibitemOpen
  \bibfield  {author} {\bibinfo {author} {\bibfnamefont {K.}~\bibnamefont {Schwarz}}\ and\ \bibinfo {author} {\bibfnamefont {P.}~\bibnamefont {Blaha}},\ }\bibfield  {title} {\enquote {\bibinfo {title} {Solid state calculations using {WIEN2k}},}\ }\href@noop {} {\bibfield  {journal} {\bibinfo  {journal} {Comput. Mater. Sci.}\ }\textbf {\bibinfo {volume} {28}},\ \bibinfo {pages} {259} (\bibinfo {year} {2003})}\BibitemShut {NoStop}%
\bibitem [{\citenamefont {Perdew}\ \emph {et~al.}(1996)\citenamefont {Perdew}, \citenamefont {Burke},\ and\ \citenamefont {Ernzerhof}}]{perdew1996generalized}%
  \BibitemOpen
  \bibfield  {author} {\bibinfo {author} {\bibfnamefont {J.~P.}\ \bibnamefont {Perdew}}, \bibinfo {author} {\bibfnamefont {K.}~\bibnamefont {Burke}}, \ and\ \bibinfo {author} {\bibfnamefont {M.}~\bibnamefont {Ernzerhof}},\ }\bibfield  {title} {\enquote {\bibinfo {title} {Generalized gradient approximation made simple},}\ }\href@noop {} {\bibfield  {journal} {\bibinfo  {journal} {Phys. Rev. Lett.}\ }\textbf {\bibinfo {volume} {77}},\ \bibinfo {pages} {3865} (\bibinfo {year} {1996})}\BibitemShut {NoStop}%
\bibitem [{\citenamefont {Rourke}\ and\ \citenamefont {Julian}(2012)}]{julian2012numerical}%
  \BibitemOpen
  \bibfield  {author} {\bibinfo {author} {\bibfnamefont {P.}~\bibnamefont {Rourke}}\ and\ \bibinfo {author} {\bibfnamefont {S.}~\bibnamefont {Julian}},\ }\bibfield  {title} {\enquote {\bibinfo {title} {Numerical extraction of de {Haas}–van {Alphen} frequencies from calculated band energies},}\ }\href {\doibase https://doi.org/10.1016/j.cpc.2011.10.015} {\bibfield  {journal} {\bibinfo  {journal} {Comput. Phys. Commun.}\ }\textbf {\bibinfo {volume} {183}},\ \bibinfo {pages} {324} (\bibinfo {year} {2012})}\BibitemShut {NoStop}%
\end{thebibliography}
\end{document}